\documentclass[aps,preprint,superscriptaddress,nofootinbib,tightenlines]{revtex4-1}
\usepackage[utf8x]{inputenc}
\usepackage{mathrsfs}
\usepackage{latexsym,bm}
\usepackage{graphicx}
\usepackage{indentfirst}
\usepackage{slashed}
\usepackage{amsmath}
\usepackage{amssymb}
\usepackage{color}
\usepackage{hyperref}
\usepackage{epsfig}
\usepackage[titletoc]{appendix}
\usepackage{multirow}%
\usepackage{rotating}
\usepackage{epstopdf}
\usepackage{extarrows}
\usepackage{tabularx}
\usepackage{tabulary}
\usepackage{soul} 
\usepackage{xcolor}
\usepackage{graphicx}
\usepackage{lipsum}

\newcommand{\be}{\begin{equation}}
\newcommand{\ee}{\end{equation}}
\newcommand{\bea}{\begin{eqnarray}}
\newcommand{\eea}{\end{eqnarray}}
\newcommand{\beas}{\begin{eqnarray*}}
\newcommand{\eeas}{\end{eqnarray*}}

\newcommand{\al}{&}
\newcommand{\lag}{\mathcal{L}}

\newcommand{\itp}{\affiliation{CAS Key Laboratory of Theoretical Physics,
            Institute of Theoretical Physics,\\ Chinese Academy of Sciences,
            Beijing 100190, China}}
\newcommand{\bonn}{\affiliation{Helmholtz-Institut f\"ur Strahlen- und
             Kernphysik and Bethe Center for Theoretical Physics,\\
             Universit\"at Bonn,  D-53115 Bonn, Germany}}
\newcommand{\fzj}{\affiliation{Institute for
           Advanced Simulation, Institut f\"ur Kernphysik and
           J\"ulich Center for Hadron Physics,\\
           Forschungszentrum J\"ulich, D-52425 J\"ulich, Germany}}
\newcommand{\ucas}{\affiliation{School of Physical Sciences,
            University of Chinese Academy of Sciences,\\
            Beijing 100049, China}}
\newcommand{\tsu}{\affiliation{Tbilisi State University, 0186 Tbilisi, Georgia}}

\begin{document}
\thispagestyle{empty}

\title{
\Large \bf
Interactions between vector mesons and dynamically generated resonances}

\author{Meng-Lin~Du}\email{du@hiskp.uni-bonn.de}\bonn

\author{Dilege~G\"ulmez}\email{guelmez@hiskp.uni-bonn.de}\bonn

\author{Feng-Kun~Guo}\email{fkguo@itp.ac.cn}\itp\ucas

\author{Ulf-G.~Mei{\ss}ner}\email{meissner@hiskp.uni-bonn.de}\bonn\itp\fzj\tsu

\author{Qian~Wang}\email{wangqian@hiskp.uni-bonn.de}\bonn

\begin{abstract}
The $\rho\rho$ interaction and the corresponding dynamically generated bound states are revisited. We demonstrate that an
improved unitarization method is necessary to study the pole structures of amplitudes 
outside the near-threshold region. In this work, 
we extend the study of the covariant $\rho\rho$ scattering in a unitarized 
chiral theory to the $S$-wave interactions for the whole vector-meson nonet. 
We demonstrate that there are unphysical left-hand cuts in
the on-shell factorization approach of the Bethe-Salpeter equation. 
This is in conflict with the correct analytic behavior and makes the so-obtained poles, 
corresponding to possible bound states or resonances,  
unreliable. To avoid this difficulty, we employ the first iterated solution of 
the $N/D$ method and investigate the 
possible dynamically generated resonances from vector-vector interactions.
A comparison with the results from the nonrelativistic calculation is provided as well. 

\end{abstract}


\maketitle

\newpage

\section{Introduction}\label{sec:intro}

At very low energies, the strong interactions among the lowest-lying pseudoscalars, 
i.e. $\pi$, $K$ and $\eta$, are successfully described by chiral perturbation 
theory (ChPT)~\cite{Weinberg:1978kz,Gasser:1983yg,Gasser:1984gg}. 
To extend such a theory to higher energies, heavier meson resonances
must be incorporated, with the light vector-meson nonet, i.e. $\rho$, 
$K^\ast$, $\omega$ and $\phi$, being the most important one. To that end, various approaches
have been proposed,  among which the most convenient scheme was suggested in Ref.~\cite{Weinberg:1968de} 
and further developed by Callan, Coleman, Wess and Zumino in their
classical works~\cite{Coleman:1969sm,Callan:1969sn}, known as the CCWZ 
formalism. In this scheme, low-energy theorems are easily built in and 
the vector mesons transform homogeneously under a nonlinear realization of the chiral 
symmetry. While the leading order effective Lagrangian for vectors are fully determined 
by the chiral symmetry, the low-energy constants (LECs) of higher orders 
are not constrained and have to be fixed by experimental measurements and lattice data in
principle. The framework for investigating the dynamics of resonances in a chiral theory has been laid out in Refs.~\cite{Ecker:1988te,Ecker:1989yg} along these lines. Recently, a chiral expansion of the masses and decay constants of those low-lying mesons was proposed
up to one-loop order in Ref.~\cite{Bavontaweepanya:2018yds}, see also Ref.~\cite{Djukanovic:2004mm,Rosell:2004mn,Bruns:2004tj}.

Alternatively, the phenomenological success of vector meson dominance and 
especially the universality of the $\rho$-couplings~\cite{Sakurai:book},  
i.e. $g_{\rho\pi\pi}\approx g_{\rho NN}$, 
have motivated the massive Yang-Mills method~\cite{Gasiorowicz:1969kn} 
and the hidden local symmetry approach~\cite{Bando:1984ej,Bando:1985rf}. 
In these treatments, low-energy theorems impose 
important constraints on the $\rho$-couplings. Based on a linear realization 
of the chiral symmetry, both the vectors $\rho$ and its chiral partners $a_1$ must 
be treated on the same footing in the massive Yang-Mills method. The chiral 
low-energy theorems are not immediately obvious at the Lagrangian level and 
are obtained after some delicate cancellations. A further complication comes from 
the presence of the $\pi a_1$ mixing, although it can be removed by an appropriate 
shift in the definition of the axial vector fields~\cite{Gasiorowicz:1969kn}. 
Based on that a nonlinear sigma model on the coset space $G/H$ is gauge equivalent 
to a linear model with $G_\text{global}\times H_\text{local}$, the $\rho$ mesons  were 
suggested as the dynamical gauge bosons of the hidden local symmetry 
$H_\text{local}$~\cite{Bando:1984ej,Bando:1985rf}, and their masses are generated 
via the Higgs mechanism. By this treatment the effective chiral Lagrangian is 
fixed up to one parameter, and the celebrated 
Kawarabayashi--Suzuki--Riazuddin--Fayyazuddin (KSRF) relations~\cite{Kawarabayashi:1966kd,Riazuddin:1966sw},
i.e. $g_\rho = 2 g_{\rho \pi\pi} F_\pi^2$ and $M_\rho^2 = 2 g_{\rho \pi\pi}^2 F_\pi^2$, follow naturally 
by an appropriate choice of the parameter. Moreover, the inclusion of the 
electromagnetic interactions precisely yields the vector meson dominance of photon 
couplings~\cite{Bando:1984ej,Bando:1985rf}. These approaches are in principle 
equivalent, and each corresponds to a different choice of the vector field, 
see, e.g., Refs.~\cite{Meissner:1987ge,Birse:1996hd}. The different choices of 
fields merely influence the off-shell behavior of the scattering amplitudes, and lead
to the same physics. Thus,  the choice of fields depends on the convenience of the corresponding 
Lagrangian for the specific calculations. 

It is now commonly accepted that some hadronic resonances are dynamically 
generated by strong hadron-hadron interactions. To describe these states, different unitarization procedures 
were proposed. A convenient and commonly used unitarization method is the on-shell 
factorization approximation of the Bethe--Salpeter equation (BSE).  While the unitarity cut 
(physical/right-hand cut) is treated nonperturbatively, the left-hand cut (dynamic 
sigularities~\cite{Badalian:1981xj}) are incorporated in a perturbative way~\cite{Oller:1997ti,Oller:2000fj}.
The coupled-channel version of the unitarization procedure was used to study the $S$-wave 
kaon-nucleon interactions for strangeness $S=-1$ channel in a modern framework in 
Ref.~\cite{Oller:2000fj} (for earlier works using different regulators, see~\cite{Kaiser:1995eg,Oset:1997it}) and provided a good reproduction of the event 
distributions in the region of the $\Lambda (1405)$, which was predicted as 
a $\bar KN$ hadronic molecule by Dalitz and Tuan long ago~\cite{Dalitz:1959dn}. More interestingly, the
$\Lambda (1405)$ is replaced by two nearby poles, leading to the so-called two-pole structure 
of the $\Lambda (1405)$~\cite{Oller:2000fj,Jido:2003cb} (for brief reviews, see Refs.~\cite{Guo:2017jvc,pdg2018}),
due to the $\pi\Sigma$ and $\bar K N$ coupled channels. The two-pole structure was confirmed by experiments later~\cite{Lu:2013nza}.
Recently, a similar two-pole nature of the $D_0^\ast$ 
was reported in Refs.~\cite{Albaladejo:2016lbb,Du:2017zvv}, which is based on 
the unitarized chiral effective theory constrained by lattice calculations~\cite{Liu:2012zya} and 
backed by the high quality experimental data collected at the LHCb experiment~\cite{Aaij:2016fma}.
In addition, the unitarized chiral method was employed to reveal the nature of the
$f_0(980)$ as a dynamically generated resonance in the isoscalar $\pi\pi$--$K\bar{K}$ 
system~\cite{Oller:1997ti}. 

The first attempt to investigate the dynamically generated resonances 
(including bound states) by $S$-wave $\rho\rho$ interactions was given in 
Ref.~\cite{Molina:2008jw} with the potentials derived in the framework of the hidden local
symmetry approach.  In that work, the nonrelativistic limit, i.e. $|{\mathbf p}|^2/M_\rho^2 \to 
0$ with ${\mathbf p}$ the three-momentum of the $\rho$, was taken. It is found that the $\rho\rho$ interactions
in the channel $I=0$, $J=0$ and 
$I=0$, $J=2$ channels, with $I$ the isospin and $J$ the total spin, are attractive enough to produce bound states, 
which are assigned to the $f_0(1370)$ and $f_2(1270)$ resonances, respectively. 
Furthermore, the fact that the tensor meson is lighter than the scalar one is attributed to the stronger 
attraction in the corresponding channel. However, as pointed out in 
Ref.~\cite{Gulmez:2016scm}, the $f_2(1270)$ fits very well within a nearly 
ideally-mixed $P$-wave $q\bar{q}$ nonet~\cite{Lichtenberg:1978pc,Koll:2000ke,Ricken:2000kf}.
The $q\bar{q}$ picture of the $f_2(1270)$ is also supported 
by the experimental data on $\gamma\gamma\to \pi\pi$~\cite{Dai:2014zta,Mori:2007bu}. One notices that the $\rho\rho$
bound state assignment of  the $f_2(1270)$ in Ref.~\cite{Molina:2008jw} is based on
the nonrelativistic limit although the binding momentum in this case reaches about 445~MeV. The interaction was over-extrapolated 
to the region where the relativistic effect has to be taken into account, see 
e.g. Ref.~\cite{Gulmez:2016scm}. Moreover, the $t$- and $u$-channel $\rho$-exchange diagrams 
shrink to contact terms for the four-$\rho$ interactions in Ref.~\cite{Molina:2008jw} and thus 
the corresponding left-hand cuts are neglected. A scrutinization of the extreme nonrelativistic 
approximation can be found in Ref.~\cite{Gulmez:2016scm}, where a 
covariant formalism using relativistic propagators is employed and possible generated resonances are revisited.
Furthermore, the unitarization procedure is improved using the first iterated solution of the $N/D$ dispersion relation.
It turns out that while a bound state pole corresponding to the $f_0(1370)$ is found, there is
no pole which can be associated with the $f_2(1270)$. 

In a following work, i.e. Ref.~\cite{Geng:2016pmf}, the authors examine the 
the relativistically covariant $\rho\rho$ interaction and argue that the disappearance of 
the tensor bound state in Ref.~\cite{Gulmez:2016scm} is due to the on-shell 
factorization of the potential done in the region where an ``unphysical'' discontinuity (imaginary part) is
developed by the left-hand cut. 
The authors of  Ref.~\cite{Geng:2016pmf} argue that the one-loop integrals for the $t$-channel $\rho$-exchange triangle and
box diagrams do not have an imaginary part and thus the left-hand cut $s\leq 3 M_\rho^2$ would be unphysical. 
The singularities of the triangle and box diagrams by putting at least two intermediate $\rho$ mesons on shell are the
triangle and box Landau singularities~\cite{Landau:1959fi}, and they indeed do not appear in the physical Riemann sheet for the
processes in the  energy region of interest according to the Coleman--Norton theorem~\cite{Coleman:1965xm}.\footnote{The left-hand
cut $s\leq 3 M_\rho^2$ develops when the Mandelstam variable $t\geq M_\rho^2$, corresponding to the fact that the $\rho$ can be
exchanged in the $t$-channel. This left-hand cut is manifest in the tree-level potential as well as in the full $T$-matrix.}
It is only due to the use of on-shell factorization so that such a left-hand cut appears wrongly on the physical Riemann sheet,
and is thus called ``unphysical''. It is worthwhile to emphasize that this issue and the related on-shell factorization problem
have been overcome in Ref.~\cite{Gulmez:2016scm} by using the first iterated solution of the $N/D$ dispersion relation, and still no tensor pole was found as mentioned above. 

At the first sight, it seems to be 
surprising that a bound state appears in the sector $(I,J)=(0,0)$, while there is no  $(I,J)=(0,2)$ bound state though the 
interaction is also attractive with a strength at threshold twice of that in the scalar sector. As is well-known, in the energy range
not very close to the threshold, higher orders in the effective range expansion, and thus the energy dependence in the potential,
become important. That is to say, in addition to the 
attractive strength at threshold, which is proportional to the scattering length $a$, the effective  range $r_0$ is also crucial to
form a deeply-bound state. The effective range is determined 
by $$T(s)=\frac{8\pi\sqrt{s}}{-1/a+r_0p^2/2-ip},$$ around the threshold 
with $p$ the three-momentum  and $s=E^2$ the total energy squared in the center-of-mass frame. It is easy to 
see that $r_0$ is proportional to the derivative of $\sqrt{s}T^{-1}(s)$ with respect to the energy 
at the threshold. As a result, with the increasing of the slope of the potential 
as a function of energy, the effective range decreases. As can be seen from Fig.~4 
in Ref.~\cite{Gulmez:2016scm}, the effective range for $(I,J)=(0,0)$ is much larger than that in the tensor sector. 
A naive calculation with the tree-level potential shows that the tensor sector even  
has a negative effective range.\footnote{However, the scalar and tensor potentials
have the same energy dependence in Ref.~\cite{Geng:2016pmf} (see Table I) and thus have the same effective range. }
From the $D$ function plotted in Fig.~11 in Ref.~\cite{Gulmez:2016scm}, where the problems of the on-shell factorization have
been cured, it can be seen that 
the disappearance of the tensor bound state stems from the energy-dependence 
of the potential instead of the left-hand cut.

The extension of the vector-vector interaction to SU(3) is given in Ref.~\cite{Geng:2008gx}. Up to 11 dynamically 
generated states were reported. While six of them were assigned to the $f_0(1370)$, 
$f_0(1710)$, $f_2(1270)$, $f_2^\prime (1525)$, $a_2(1320)$ and $K_2^\ast (1430)$ 
resonances, more states with quantum numbers of the $h_1$, $a_0$, 
$b_1$, $K_0^\ast$ and $K_1$ were predicted. However, 
since that work employed the same nonrelativistic formalism as that in Ref.~\cite{Molina:2008jw}, a recalculation
keeping the $t$- and $u$-channel vector-meson propagators using the method of Ref.~\cite{Gulmez:2016scm} is needed,
which is the scope of this paper.

This work is organized as follows. In section~\ref{sec:formalsim}, the effective 
Lagrangian in the hidden local symmetry approach is briefly introduced and 
the scattering amplitudes are calculated, followed by the partial wave projection. In section~\ref{sec:bse}, 
the on-shell factorization of the BSE is applied to the single-channels, and poles are searched for in the energy range outside of 
the left-hand cuts. The improved unitarization formula using the first iterated 
solution of the $N/D$ dispersion relation~\cite{Gulmez:2016scm} is employed in section~\ref{sec:nd}. Finally, 
section~\ref{sec:summary} comprises a summary and outlook.

\section{Formalism}\label{sec:formalsim}

\subsection{Effective Lagrangian in hidden local symmetry approach}\label{subsec:lag}

Various approaches including the light vector mesons in an effective theory 
respecting the chiral symmetry are proven to be equivalent, e.g. see Refs.~\cite{Meissner:1987ge,Birse:1996hd}. In this work,
we employ the hidden local symmetry  formalism in which the vectors are treated as gauge bosons of a hidden local 
symmetry transforming inhomogeneously. In this approach,
the phenomenologically successful KSRF relations, vector-meson dominance and the universality of $\rho$-couplings,
as well as the Weinberg--Tomozawa theorem for $\pi$-$\rho$ scattering, can be  obtained.

In the hidden local symmetry approach, the global chiral symmetry is encoded into 
a SU(3) matrix $U(x)$. The local symmetry is introduced by factorizing $U(x)$ 
into two SU(3) matrices~\cite{Bando:1984ej,Meissner:1987ge,Birse:1996hd}
\bea
U(x)=\xi_L ^\dag(x) \xi_R^{}(x).
\eea
This factorization is arbitrary at each space-time point, which is equivalent to a 
local SU(3) symmetry. Vector mesons $V_\mu$ are introduced as the gauge bosons of this local symmetry.
The  leading order Lagrangian has the form~\cite{Bando:1984ej,Meissner:1987ge,Birse:1996hd}
\bea\label{lag:hg}
\lag = \frac{F_\pi^2}{4}\langle (L_\mu -R_\mu)^2\rangle + a \frac{F_\pi^2}{4}
\langle (L_\mu+R_\mu)^2\rangle -\frac{1}{4}\langle V_{\mu\nu}V^{\mu\nu}\rangle,
\eea
where $F_\pi$ is the pion decay constant, $F_\pi = 93$ MeV~\footnote{Note that we use an older value here for a
better comparison with the earlier results.}, $\langle \dots\rangle$ represents a trace over SU(3) flavor space,
and $a$ is a real parameter. Further,
\bea\label{lag:def:LR}
R_\mu \al = \al -i \big[ (\partial_\mu \xi_L)\xi^\dag_L - ig V_\mu \big], 
\nonumber\\
L_\mu \al = \al -i \big[ (\partial_\mu \xi_R)\xi_R^\dag - ig V_\mu \big], 
\eea
and 
\bea\label{lag:def:FS}
V_{\mu\nu}= \partial_\mu V_\nu -\partial_\nu V_\mu -ig [V_\mu, V_\nu],
\eea
is the corresponding gauge-covariant field strengths with $g$ the gauge coupling constant. 

By choosing the unitary gauge, i.e. 
\bea\label{eq:unigauge}
\xi_R(x)=\xi_L^\dag (x) =u (x),
\eea
one obtains the Lagrangian 
\bea\label{lag:uni}
\lag = \frac{F_\pi^2}{4}\langle u_\mu u^\mu\rangle +a F_\pi^2 \langle (i\Gamma_\mu 
-gV_\mu )^2\rangle -\frac{1}{4}\langle V_{\mu\nu}V^{\mu\nu}\rangle,
\eea
where 
\bea
u_\mu \al = \al i \big( u^\dag \partial_\mu u - u \partial_\mu u^\dag \big), 
\nonumber \\
\Gamma_\mu \al = \al \frac12 \big( u^\dag \partial u + u \partial_\mu u^\dag\big).
\eea
The Goldstone bosons $\Phi$ are nonlinearly encoded in $u(x) = \exp \big( i\Phi/
(\sqrt{2}F_\pi) \big)$ with
\bea
\Phi = 
\begin{pmatrix}
\frac{1}{\sqrt{2}}\pi^0 + \frac{1}{\sqrt{6}}\eta &  \pi^+ & K^+ \\
\pi^- & -\frac{1}{\sqrt{2}}\pi^0+\frac{1}{\sqrt{6}}\eta & K^0 \\
K^- & \bar{K}^0 & -\frac{2}{\sqrt{6}}\eta
\end{pmatrix}, 
\eea
and the vector-meson matrix $V_\mu$ has the form 
\bea
V_\mu = \begin{pmatrix}
\frac{1}{\sqrt{2}}\rho^0+\frac{1}{\sqrt{2}}\omega & \rho^+ & K^{\ast +} \\
\rho^- & -\frac{1}{\sqrt{2}}\rho^0+\frac{1}{\sqrt{2}}\omega & K^{\ast 0}\\
K^{\ast -} & \bar{K}^{\ast 0} & \phi
\end{pmatrix}_\mu .
\eea

The first term in Eq.~\eqref{lag:uni} is identical to the familiar leading order chiral 
Lagrangian for pseudoscalar mesons~\cite{Gasser:1984gg},  the second term 
contains the $V\Phi\Phi$ interaction and the vector mass term, and the kinetic 
term (as well as the self-interaction) of $V_\mu$ is given in the last term. 
Expanding the Lagrangian~\eqref{lag:uni} up to two pseudoscalars, one 
finds
\bea\label{lag:ksrf}
\lag_\text{mass~} \al = \al a g^2 F_\pi^2 \langle V_\mu V^\mu\rangle,\nonumber \\
\lag_{V\Phi\Phi~} \al = \al -i \frac{a}{2}g\langle V^\mu [\Phi, \partial_\mu \Phi] 
\rangle, \nonumber \\
\lag_{VVV~} \al = \al ig \langle (\partial_\mu V_\nu -\partial_\nu V_\mu )V^\mu 
V^\nu \rangle , \nonumber \\
\lag_{VVVV}\al = \al \frac{1}{2}g^2 \langle V_\mu V_\nu V^\mu V^\nu 
- V_\mu V^\mu V_\nu V^\nu \rangle .
\eea
In this work, we choose $a=2$, which leads to the celebrated KSRF 
relations and universality of the $\rho$ couplings.

\subsection{Scattering amplitudes}\label{subsec:amp}

\begin{figure}[htbp]
\vspace{8mm}
\begin{center}
\includegraphics[width=0.5\textwidth]{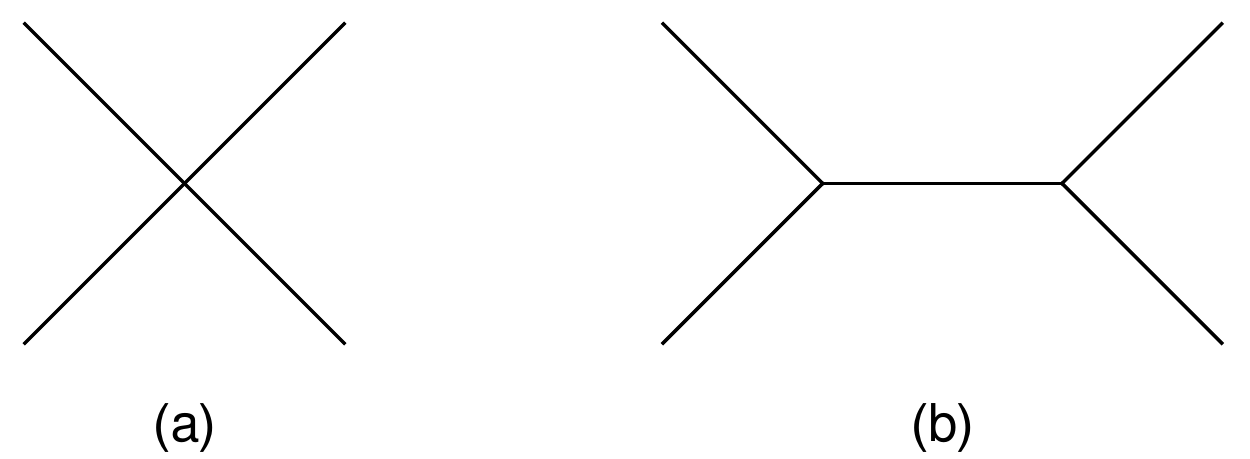}
\end{center}
\vspace{-3mm}
\caption{Tree-level Feynman diagrams for the 
vector-vector scattering. The $u$- and $t$-channel vector-exchange diagrams 
are not shown explicitly. }\label{fig:feyngraph}
\end{figure}

With the Lagrangian given in Eq.~\eqref{lag:ksrf}, we are in the position to calculate the 
vector-vector scattering amplitudes. At tree level, the Feynman diagrams needed 
are displayed in Fig.~\ref{fig:feyngraph}, 
where the $t$- and $u$-channel 
vector-exchange diagrams are not shown explicitly. The amplitude for the process 
$V_1(p_1)V_2(p_2)\to V_3(p_3)V_4(p_4)$ can be written as 
\bea\label{eq:ampform}
A(s,t,u) \al = \al A_c(s,t,u) + A_s(s,t,u) +A_t(s,t,u) +A_u(s,t,u),
\eea
where $A_c$, $A_s$, $A_t$ and $A_u$ correspond to the four-vector contact diagrams, i.e. (a) in Fig.~\ref{fig:feyngraph}, $s$-, $t$- and $u$-channel 
vector-exchange diagrams, respectively. Here, the Mandelstam variables are 
defined as $s=(p_1+p_2)^2$, $t=(p_1-p_3)^2$ and $u=(p_1-p_4)^2$, which satisfy 
the constraint $s+t+u=\sum_i p_i^2 =\sum_i M_{V_i}^2$, where the last equality holds when the initial and final vector mesons are on shell. 

The contributions from the contact amplitudes from $\lag_{VVVV}$ do not depend on the vector momenta explicitly, and are given by 
\bea\label{eq:amp:cont}
A_c(s,t,u) \al =\al  g^2 C_1 \left( 2\epsilon_1\cdot
 \epsilon_2 ~\epsilon^\ast_3\cdot 
\epsilon^\ast_4 -\epsilon_1\cdot\epsilon^\ast_3 ~\epsilon_2\cdot\epsilon_4^\ast 
-\epsilon_1\cdot\epsilon_4^\ast ~\epsilon_2\cdot\epsilon_3^\ast \right)\nonumber\\
\al \al  - g^2 C_2 \left(2\epsilon_1\cdot\epsilon_4^\ast ~\epsilon_2\cdot\epsilon_3^\ast -\epsilon_1\cdot\epsilon_2~\epsilon_3^\ast \cdot\epsilon_4^\ast 
-\epsilon_1\cdot\epsilon_3^\ast~\epsilon_2\cdot\epsilon_4^\ast \right)
\nonumber \\
\al \equiv \al C_1 A_c^1 + C_2 A_c^2 ,
\eea
where $\epsilon_i^{(\ast)}$ is the polarization vector of the $i^\text{th}$ 
external vector meson. The $C_i$'s are coupling constants given in
Table~\ref{tab:amps}.
The polarization vector can be characterized by the 
corresponding three-momentum ${\mathbf p}_i$ and the third component of the spin in 
its rest frame, and the explicit expression for the polarization vectors 
can be found in Appendix A of Ref.~\cite{Gulmez:2016scm}. The two 
structures in Eq.~\eqref{eq:amp:cont} are symmetric by exchanging 
$\epsilon_2\leftrightarrow \epsilon_4^\ast$ due to the two operators in 
$\lag_{VVVV}$, c.f. Eq.~\eqref{lag:def:FS}. 

Considering the vector-exchange amplitudes, for the $s$-channel diagram exchanging 
a vector $V$ with mass $M_V$, it takes the form 
\bea
A(s,t,u) \al = \al C_s^V A_s^V (s,t,u)  \\
\al \equiv \al  C_s^V \frac{g^2}{s-M_V^2}\bigg[ \epsilon_1\cdot\epsilon_2 ~
\epsilon_3^\ast\cdot\epsilon_4^\ast ~\Big(u-t-\frac{(M_1^2-M_2^2)(M_3^2-M_4^2)}{M_V^2}\Big)  \nonumber \\
\al +\al 4 \Big( \epsilon_3^\ast\cdot \epsilon_4^\ast~ 
(\epsilon_1\cdot p_2~\epsilon_2\cdot p_3 - \epsilon_1\cdot p_3~\epsilon_2\cdot 
p_1) +\epsilon_2\cdot p_1~ (\epsilon_4^\ast\cdot p_3~\epsilon_1\cdot \epsilon_3^\ast
- \epsilon_3^\ast\cdot p_4 ~\epsilon_1\cdot\epsilon_4^\ast ) \nonumber \\
\al + \al 
\epsilon_1\cdot p_2 (\epsilon_3^\ast\cdot p_4~\epsilon_2\cdot \epsilon_4^\ast 
-\epsilon_4^\ast\cdot p_3~\epsilon_2\cdot \epsilon_3^\ast ) \Big) 
-4\epsilon_1\cdot\epsilon_2~ (
\epsilon_3^\ast\cdot p_1~\epsilon_4^\ast\cdot p_3 - \epsilon_3^\ast\cdot p_4~
\epsilon_4^\ast\cdot p_1 )  \bigg], \nonumber 
\eea
with $M_i$ the mass of the $i^\text{th}$ external vector meson. The $t$- and 
$u$-channel amplitudes $A_t^V$ and $A_u^V$ can be obtained from $A_s^V$ by 
performing the exchange $p_2 \leftrightarrow -p_3$, $\epsilon_2\leftrightarrow
\epsilon_3^\ast$ and $p_2 \leftrightarrow -p_4$, $\epsilon_2\leftrightarrow
\epsilon_4^\ast$, respectively.  $C_s^V$ is a coupling constant given in
Table~\ref{tab:amps}.

It is more convenient to study the scattering amplitudes in the isospin basis 
instead of the particle basis. The scattering processes can be classified by 
the strangeness $S$ and isospin $I$ of the system. There are 7 independent 
combinations in total. The corresponding quantum numbers $(S,I)$ of the scattering 
systems are $(0,0)$, $(0,1)$, $(0,2)$, $(1,1/2)$, $(1,3/2)$, $(2,0)$ and 
$(2,1)$, among which $(0,2)$, $(1,3/2)$, $(2,0)$ and $(2,1)$ are single-channel processes. 
For $(S,I)=(0,0)$, there are five channels: $\rho\rho$, $K^\ast \bar{K}^\ast$, 
$\omega\omega$, $\omega\phi$ and $\phi\phi$. For $(S,I)=(0,1)$, there are four channels:
$\rho\rho$, $K\bar{K}^\ast$, $\rho\omega$ and $\rho\phi$, 
and for $(S,I)=(1,1/2)$ there are three channels: $\rho K^\ast$, $K^\ast\omega$ and $K^\ast\phi$. 
The phase convention we use to relate the particle basis to the isospin basis is such that
\bea
\left| \rho^+\right\rangle = -\left| 1,+1\right\rangle, \quad 
\left| \bar{K^\ast}^0 \right\rangle =-\left| \frac12, +\frac12 \right\rangle,
\eea
while all the other states have a positive sign.

In the $(S,I)$ basis, the tree level scattering amplitudes are given by 
\bea\label{eq:amps}
A^{(S,I)} (s,t,u) = \sum_{i=1,2} C_i^{(S,I)}A_c^i (s,t,u) +\sum_{\substack{V
=\rho, K^\ast, \omega,\phi \\ j=s,t,u}} C_j^{V(S,I)}
A_j^V(s,t,u)  ,
\eea
where the coefficients are collected in Table~\ref{tab:amps}.

\begin{table}[htbp]
\caption{Coefficients of the tree-level vector-vector scattering amplitudes in the
$(S,I)$ basis. }\label{tab:amps}
\vspace{-0.5cm}
\bea
\begin{array}{|ll|cccccccccccccc|}
\hline\hline
(S,I) & \text{Channel} & C_1 & C_2 & C_t^\rho & C_t^{K^\ast} & C_t^\omega &
C_t^\phi &  C_u^\rho & C_u^{K^\ast} & C_u^\omega & C_u^\phi &  
C_s^\rho & C_s^{K^\ast} & C_s^\omega & C_s^\phi \\ 
\hline
(0,0) & \rho \rho \to \rho \rho  & 4 & 0 & -4 & 0 & 0 & 0 & -4 & 0 & 0 & 0 & 0 & 0 & 0 & 0 \\
\, & \rho \rho \to \text{{\scriptsize $K^* \bar{K}^*$}} & \sqrt{\frac{3}{2}} & 0 & 0 & -\sqrt{\frac{3}{2}} & 0 & 0 & 0 & -\sqrt{\frac{3}{2}} & 0 &
   0 & 0 & 0 & 0 & 0 \\
\, & \rho \rho \to \omega \omega  & 0 & 0 & 0 & 0 & 0 & 0 & 0 & 0 & 0 & 0 & 0 & 0 & 0 & 0 \\
\, & \rho \rho \to \omega \phi  & 0 & 0 & 0 & 0 & 0 & 0 & 0 & 0 & 0 & 0 & 0 & 0 & 0 & 0 \\
\, & \rho \rho \to \phi \phi  & 0 & 0 & 0 & 0 & 0 & 0 & 0 & 0 & 0 & 0 & 0 & 0 & 0 & 0 \\
\, & \text{{\scriptsize $K^* \bar{K}^*$}} \to \text{{\scriptsize $K^* \bar{K}^*$}} & 0 & 3 & -\frac{3}{2} & 0 & -\frac{1}{2} & -1 & 0 & 0 & 0 & 0 & 0 & 0 & -1 & -2 \\
\, & \text{{\scriptsize $K^* \bar{K}^*$}}\to \omega \omega  & -\frac{1}{\sqrt{2}} & 0 & 0 & \frac{1}{\sqrt{2}} & 0 & 0 & 0 & \frac{1}{\sqrt{2}} &
   0 & 0 & 0 & 0 & 0 & 0 \\
\, & \text{{\scriptsize $K^* \bar{K}^*$}}\to \omega \phi  & 1 & 0 & 0 & -1 & 0 & 0 & 0 & -1 & 0 & 0 & 0 & 0 & 0 & 0 \\
\, & \text{{\scriptsize $K^* \bar{K}^*$}}\to \phi \phi  & -\sqrt{2} & 0 & 0 & \sqrt{2} & 0 & 0 & 0 & \sqrt{2} & 0 & 0 & 0 & 0 & 0 & 0 \\
\, & \omega \omega \to \omega \omega  & 0 & 0 & 0 & 0 & 0 & 0 & 0 & 0 & 0 & 0 & 0 & 0 & 0 & 0 \\
\, & \omega \omega \to \omega \phi  & 0 & 0 & 0 & 0 & 0 & 0 & 0 & 0 & 0 & 0 & 0 & 0 & 0 & 0 \\
\, & \omega \omega \to \phi \phi  & 0 & 0 & 0 & 0 & 0 & 0 & 0 & 0 & 0 & 0 & 0 & 0 & 0 & 0 \\
\, & \omega \phi \to \omega \phi  & 0 & 0 & 0 & 0 & 0 & 0 & 0 & 0 & 0 & 0 & 0 & 0 & 0 & 0 \\
\, & \omega \phi \to \phi \phi  & 0 & 0 & 0 & 0 & 0 & 0 & 0 & 0 & 0 & 0 & 0 & 0 & 0 & 0 \\
\, & \phi \phi \to \phi \phi  & 0 & 0 & 0 & 0 & 0 & 0 & 0 & 0 & 0 & 0 & 0 & 0 & 0 & 0 \\
(0,1)  & \rho \rho \to \rho \rho  & -2 & 4 & -2 & 0 & 0 & 0 & 2 & 0 & 0 & 0 & -4 & 0 & 0 & 0 \\
\, & \rho \rho \to \text{{\scriptsize $K^* \bar{K}^*$}}& -1 & 2 & 0 & -1 & 0 & 0 & 0 & 1 & 0 & 0 & -2 & 0 & 0 & 0 \\
\, & \rho \rho \to \rho \omega  & 0 & 0 & 0 & 0 & 0 & 0 & 0 & 0 & 0 & 0 & 0 & 0 & 0 & 0 \\
\, & \rho \rho \to \rho \phi  & 0 & 0 & 0 & 0 & 0 & 0 & 0 & 0 & 0 & 0 & 0 & 0 & 0 & 0 \\
\, & \text{{\scriptsize $K^* \bar{K}^*$}}\to \text{{\scriptsize $K^* \bar{K}^*$}} & 0 & 1 & \frac{1}{2} & 0 & -\frac{1}{2} & -1 & 0 & 0 & 0 & 0 & -1 & 0 & 0 & 0 \\
\, & \text{{\scriptsize $K^* \bar{K}^*$}}\to \rho \omega  & \frac{1}{\sqrt{2}} & 0 & 0 & -\frac{1}{\sqrt{2}} & 0 & 0 & 0 & -\frac{1}{\sqrt{2}} & 0
   & 0 & 0 & 0 & 0 & 0 \\
\, & \text{{\scriptsize $K^* \bar{K}^*$}}\to \rho \phi  & -1 & 0 & 0 & 1 & 0 & 0 & 0 & 1 & 0 & 0 & 0 & 0 & 0 & 0 \\
\, & \rho \omega \to \rho \omega  & 0 & 0 & 0 & 0 & 0 & 0 & 0 & 0 & 0 & 0 & 0 & 0 & 0 & 0 \\
\, & \rho \omega \to \rho \phi  & 0 & 0 & 0 & 0 & 0 & 0 & 0 & 0 & 0 & 0 & 0 & 0 & 0 & 0 \\
\, & \rho \phi \to \rho \phi  & 0 & 0 & 0 & 0 & 0 & 0 & 0 & 0 & 0 & 0 & 0 & 0 & 0 & 0 \\
(0,2) & \rho \rho \to \rho \rho  & -2 & 0 & 2 & 0 & 0 & 0 & 2 & 0 & 0 & 0 & 0 & 0 & 0 & 0 \\
(1,\frac12 )& \rho \text{{\scriptsize $K^\ast$}}\to  \rho\text{{\scriptsize $K^*$}}  & \frac{1}{2} & \frac{3}{2} & -2 & 0 & 0 & 0 & 0 & -\frac{1}{2} & 0 & 0 & 0 & -\frac{3}{2} & 0
   & 0 \\
\, & \rho\text{{\scriptsize $K^*$}} \to\text{{\scriptsize $K^*$}} \omega  & 0 & -\frac{\sqrt{3}}{2} & 0 & \frac{\sqrt{3}}{2} & 0 & 0 & 0 & 0 & 0 & 0 & 0 &
   \frac{\sqrt{3}}{2} & 0 & 0 \\
\, & \rho\text{{\scriptsize $K^*$}} \to \text{{\scriptsize $K^*$}} \phi  & 0 & \sqrt{\frac{3}{2}} & 0 & -\sqrt{\frac{3}{2}} & 0 & 0 & 0 & 0 & 0 & 0 & 0 &
   -\sqrt{\frac{3}{2}} & 0 & 0 \\
\, & \text{{\scriptsize $K^*$}} \omega \to \text{{\scriptsize $K^*$}} \omega  & -\frac{1}{2} & \frac{1}{2} & 0 & 0 & 0 & 0 & 0 & \frac{1}{2} & 0 & 0 & 0 & -\frac{1}{2} &
   0 & 0 \\
\, & \text{{\scriptsize $K^*$}} \omega \to \text{{\scriptsize $K^*$}} \phi  & \frac{1}{\sqrt{2}} & -\frac{1}{\sqrt{2}} & 0 & 0 & 0 & 0 & 0 & -\frac{1}{\sqrt{2}} & 0 & 0
   & 0 & \frac{1}{\sqrt{2}} & 0 & 0 \\
\, & \text{{\scriptsize $K^*$}} \phi \to \text{{\scriptsize $K^*$}} \phi  & -1 & 1 & 0 & 0 & 0 & 0 & 0 & 1 & 0 & 0 & 0 & -1 & 0 & 0 \\
(1,\frac32 ) &\rho\text{{\scriptsize $K^*$}} \to\rho\text{{\scriptsize $K^*$}}  & -1 & 0 & 1 & 0 & 0 & 0 & 0 & 1 & 0 & 0 & 0 & 0 & 0 & 0 \\
(2,0) & \text{{\scriptsize $K^*$}}\text{{\scriptsize $K^*$}}\to \text{{\scriptsize $K^*$}}\text{{\scriptsize $K^*$}} & 0 & 0 & -\frac{3}{2} & 0 & \frac{1}{2} & 1 & \frac{3}{2} & 0 & -\frac{1}{2}
   & -1 & 0 & 0 & 0 & 0 \\
(2,1) & \text{{\scriptsize $K^*$}}\text{{\scriptsize $K^*$}}\to \text{{\scriptsize $K^*$}}\text{{\scriptsize $K^*$}} & -2 & 0 & \frac{1}{2} & 0 & \frac{1}{2} & 1 & \frac{1}{2} & 0 & \frac{1}{2}
   & 1 & 0 & 0 & 0 & 0 \\
\hline\hline
\end{array}\nonumber
\eea
\end{table}

\subsection{Partial wave amplitudes}\label{subsec:pwa}

For a relativistic system, the orbital angular momentum $L$ and the total 
spin $\Sigma$  are not good quantum numbers. Instead, the total angular momentum $J$ is conserved. 
Since we are interested in 
systems with definite strangeness and isospin $(S,I)$, hereafter we will 
neglect the $(S,I)$ index for brevity. A complete basis for the angular momentum coupling
of a two-vector system could be chosen as the total angular momentum $J$, the 
corresponding third component $M$, the orbital angular momentum $L$ and the 
total spin $\Sigma$, among which only $J$ and $M$ are good quantum numbers. 
A state $|JM,L \Sigma\rangle$ can be expressed in terms of the states 
$|{\mathbf p},\sigma_1\sigma_2\rangle$ which is the direct product of the one-particle 
states $|{\mathbf p},\sigma_1\rangle$ and $|-{\mathbf p},\sigma_2\rangle$ with $\sigma_i$ the 
third component of spin for particle $i$ and $\mathbf{p}$ the momentum in the center-of-mass frame~\cite{Gulmez:2016scm},
\bea\label{eq:spindecom}
|JM,L \Sigma\rangle = \frac{1}{\sqrt{4\pi}}\sum_{\substack{\sigma_1,
\sigma_2 \\ \Sigma_3,L_3 } } \int d \Omega_{\bold p} Y_L^{L_3} (\Omega_{\mathbf p}) 
(\sigma_1 \sigma_2 \Sigma_3|s_1 s_2\Sigma ) (L_3 \Sigma_3 M|L\Sigma J) 
|{\mathbf p},\sigma_1 \sigma_2\rangle,
\eea
with $s_i=1$  the spin of the $i^\text{th}$ vector meson, 
and $(m_1m_2 M|l_1l_2 L)$ denotes the pertinent Clebsch--Gordan coefficient. If the two vector mesons are identical particles,
an extra symmetrization factor $1/\sqrt{2}$ is required in Eq.~\eqref{eq:spindecom}. 

For a complete consideration of coupled channels, 
in general, all transitions between states with the same $J$ need to be taken into 
account with the transition matrix elements
\bea\label{eq:scattpartial}
T^{J}_{L\Sigma;L^\prime\Sigma^\prime} = \langle J M,L\Sigma|\hat{T}|
 J M, L^\prime\Sigma^\prime \rangle ,
\eea
with $\hat{T}$ the scattering operator. Note that the partial wave amplitudes in Eq.~\eqref{eq:scattpartial} are independent
of the third component $M$ due to the rotational symmetry. Thus, the expression of Eq.~\eqref{eq:scattpartial} can be
written as~\cite{Gulmez:2016scm} 
\bea\label{eq:pwp}
T^{J}_{L\Sigma;L^\prime\Sigma^\prime} \al = \al \frac{1}{2(2J+2)}Y_{L^\prime}
^0(\hat{\bold z})\sum_{\substack{\sigma_1,\sigma_2 \\ \sigma_1^\prime,\sigma_2
^\prime \\ L_3 }}\int d \Omega_{\bold p^\prime}Y_L^{L_3}(\Omega_{\bold p^\prime})^\ast (\sigma_1\sigma_2 \Sigma_3|11S)(L_3\Sigma_3\Sigma_3^\prime|L\Sigma 
J) \nonumber \\ \al \al \times  (0\Sigma_3\Sigma_3| L_3^\prime \Sigma^\prime J) 
T(p_1,p_2,p_3,p_4,\epsilon_1,\epsilon_2,\epsilon_3,\epsilon_4),
\eea
with the  $z$-axis defined as the direction of ${\bold p}_1$, i.e., ${\bold p_1}=|{\bold p}|\hat{\bold z}$, ${\bold p_2}=
-|{\bold p}|\hat{\bold z}$, ${\bold p_3}= {\bold p^\prime}$, ${\bold p_4}=-{\bold p^\prime}$, $\Sigma_3=\sigma_1+\sigma_2$ 
and $\Sigma_3^\prime = \sigma_1^\prime+\sigma_2^\prime$. In this work, we 
are only interested in the $S$-wave scattering amplitudes, thus we set $L=L^\prime=0$. 
In principle, the transition amplitudes with different orbital angular momenta  should be taken into account as
long as the total angular momentum $J$ is  conserved. However, it would introduce extreme complications due
to the large number of different  possible orbital angular momenta although we are only interested in $J=0$, 1, 
and 2. Both $S$ and $D$ waves are considered for SU(2) case 
in Ref.~\cite{Gulmez:2016scm}, where the inclusion of $S$--$D$-wave coupled channels leads to 
the presence of an artificial pole with anomalous properties. It has a negative 
residue, which is at odds with the probabilistic 
interpretation of a bound state. The presence of this extra pole is attributed 
to the unitarization which treats the left-hand cut perturbatively. Apart from 
that, including $S$--$D$-wave coupling produces a physical pole 
with properties close to that in the purely $S$-wave case. For simplicity, we only focus on the channels with 
$L=0$ and neglect contributions from higher $L$ in what follows.

The partial wave  projection Eq.~\eqref{eq:pwp} for a $t$-channel exchange amplitude would  develop a left-hand cut by 
\bea\label{eq:pwpt}
\frac{1}{2} \al \al \!\!\!\!\! \int_{-1}^{+1} d\cos  \theta  \frac{1}{t-M^2+i\epsilon} 
 = -\frac{s}{\sqrt{\lambda(s,M_1^2,M_2^2)\lambda(s,M_3^2,M_4^2)}} \nonumber\\
\al\times \al \log \frac{M_1^2+M_3^2-\frac{(s+M_1^2-M_2^2)(s+M_3^2-M_4^2)}{2s}-
\frac{\sqrt{\lambda(s,M_1^2,M_2^2)\lambda(s,M_3^2,M_4^2)}}{2s}-M^2+i\epsilon}{M_1^2+M_3^2-\frac{(s+M_1^2-M_2^2)(s+M_3^2-M_4^2)}{2s}+
\frac{\sqrt{\lambda(s,M_1^2,M_2^2)\lambda(s,M_3^2,M_4^2)}}{2s}-M^2+i\epsilon},
\eea
with $\lambda(a,b,c)=a^2+b^2+c^2-2ab-2bc-2ac$ the K\"all\'en function. A 
projection for a $u$-channel can be obtained from Eq.~\eqref{eq:pwpt} by 
the interchanging $p_3\leftrightarrow p_4$. 
One notices that the location of the branch points depend on the involved masses, thus on the channel, and $M_i^2$ in the above equation
comes from putting the external particles on their mass shells, $p_i^2=M_i^2$. In the coupled-channel case, the intermediate particles are
not on shell. If they are put on shell, then the corresponding left-hand cuts would be transported into all coupled channels via
Eq.~\eqref{eq:bse}, and become unphysical for the $T$-matrix elements for which they are not the external particles. 
In general, the presence of such unphysical left-hand cuts is not 
disturbing as long as they are located far below the lowest threshold.

\section{On-shell factorization of the Bethe--Salpeter equation}\label{sec:bse}

The  on-shell unitarization approach has been applied with great phenomenological success in 
previous works, see e.g.~\cite{Oller:1997ti,Oller:2000fj,Jido:2003cb,Guo:2012yt,Liu:2012zya,Albaladejo:2016lbb,Inoue:2001ip,GarciaRecio:2002td,Hyodo:2007jq,Guo:2017jvc} (and references therein),
despite of the presence of the unphysical left-hand cuts due to coupled channels. 
In these cases, the unphysical left-hand cuts are far away from the energy regions of interest, and thus do not cause serious problems.
Nevertheless, it is known that the unphysical left-hand cuts can lead to bad analytic behavior of the coupled-channel amplitudes, see, e.g., Ref.~\cite{Du:2017ttu}.
In this section, we focus on the unitarization using the on-shell factorization, and expose the problems of this method explicitly.
The basic equation for  the unitarized amplitude $T$ is given in matrix form by 
\bea\label{eq:bse}
T^J(s) = V^J(s)\cdot \big[ 1 - G(s)\cdot V^J(s)\big]^{-1}, 
\eea
where $V^J$ denotes the partial-wave projected amplitudes, and $G(s)$ is 
a diagonal matrix $G(s)=\text{diag}\{g_i(s)\}$ with $i$ the channel index. 
The fundamental loop integral $g_i(s)$ reads 
\bea
g_i(s)=i\int \frac{d^4q}{(2\pi)^4}\frac{1}{(q^2-M_1^2+i\epsilon)((P-q)^2 
-M_2^2+i\epsilon)},
\eea
with $P^2=s$ and $M_{1,2}$ the masses of the particles in that channel. The $T$-matrix has poles at the zeros of the
denominator,
\begin{equation}
D_U(s)=1-G(s) V(s)~,
\end{equation}
or its determinant, i.e., 
\begin{equation}
\text{Det}\equiv \text{det}[1 - G(s)\cdot V^J(s)]~,
\label{eq:det}
\end{equation}
for the single-channel or coupled-channel cases, respectively. 

The above loop integral is logarithmically divergent and can be 
calculated with a once-subtracted dispersion relation. In this way, 
its explicit expression is \cite{Oller:1998zr,Oller:2000fj}
\bea\label{eq:ga}
g_i(s)& =& \dfrac{1}{16\pi^2}\Big( a(\mu)+\log\frac{M_1^2}{\mu^2} 
+\frac{s-M_1^2+M_2^2}{2s}\log\frac{M_2^2}{M_1^2} +\frac{\sigma}{2s}\big[ 
\log (s-M_2^2+M_1^2+\sigma)  \\
& & -\log(-s+M_2^2-M_1^2+\sigma) +\log(s+M_2^2-M_1^2+\sigma) 
-\log(-s-M_2^2+M_1^2+\sigma) \big] \Big), \nonumber
\eea
with 
\bea\label{eq:sigma}
\sigma = \sqrt{\big(s-(M_1+M_2)^2\big)\big(s-(M_1-M_2)^2\big)}
\eea
and $a(\mu)$ is a renormalization scale ($\mu$) dependent subtraction constant.

Alternatively, the loop integral 
can be calculated with a sharp momentum-cutoff ($q_\text{max}$) regularization. The explicit expression of the cutoff-regularized loop integral, 
denoted as $g_i^c(s)$, reads~\cite{Oller:1998hw,Guo:2005wp}
\bea\label{eq:gcut}
g_i^c(s) &=& \frac{1}{32\pi^2}\Bigg\{ -\frac{\Delta}{s}\log\frac{M_1^2}{M_2^2} 
+\frac{\sigma}{s} \bigg[ \log\frac{s-\Delta+\sigma\sqrt{1+\frac{M_1^2}{q_\text{max}^2}}}{-s+\Delta+\sigma\sqrt{1+\frac{M_1^2}{q_\text{max}^2}}}
+\log \frac{s + \Delta+\sigma\sqrt{1+\frac{M_2^2}{q_\text{max}^2}}}{-s-\Delta+\sigma\sqrt{1+\frac{M_2^2}{q_\text{max}^2}}} \bigg] \nonumber \\
&+&2\frac{\Delta}{s}\log\frac{1+\sqrt{1+\frac{M_1^2}{q_\text{max}^2}}}{1+\sqrt{1+\frac{M_2^2}{q_\text{max}^2}}} -2\log\Bigg[ \Bigg( 1+\sqrt{1+\frac{M_1^2}{q_\text{max}^2}}\, \Bigg) \Bigg( 1+\sqrt{1+\frac{M_2^2}{q_\text{max}^2}}\,\Bigg)\Bigg] 
+\log\frac{M_1^2M_2^2}{q_\text{max}^4} \Bigg\}, \nonumber\\
\eea
with $\Delta = M_2^2-M_1^2$. Without any prior knowledge of the subtraction 
constant $a(\mu)$ (although its natural size is known~\cite{Oller:2000fj}), the cutoff regularization 
enables us to evaluate the loop function $g^c_i(s)$ with a natural value for the cutoff
$q_\text{max}\sim M_V$.  In this work, to investigate the stability of the generated poles to the cutoff, 
we employ the values $q_\text{max}=0.775$~GeV, 0.875~GeV and 1.0~GeV 
successively as in previous works~\cite{Molina:2008jw,Geng:2008gx,Gulmez:2016scm}.

\begin{figure}[tbh]
\vspace{8mm}
\begin{center}
\includegraphics[width=1.0\textwidth]{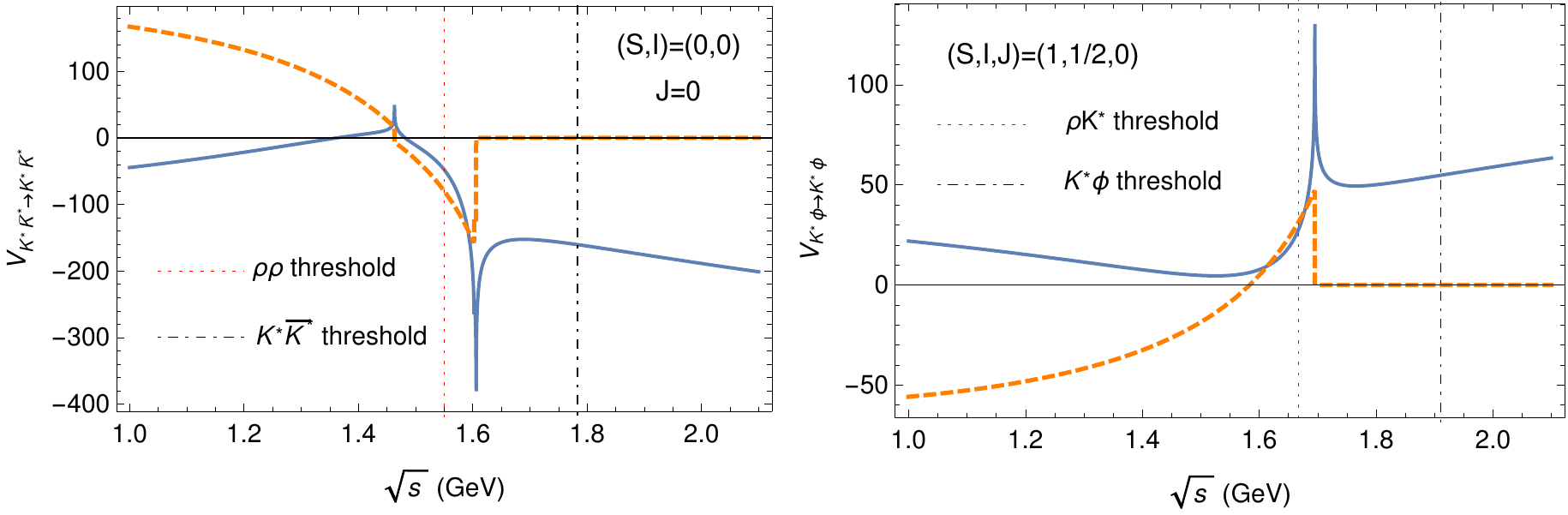}
\end{center}
\vspace{-3mm}
\caption{$S$-wave potentials for $K^\ast\bar{K}^\ast\to K^\ast\bar{K}^\ast$ 
with $(S,I,J)=(0,0,0)$ and $K^\ast\phi\to K^\ast\phi$ with $(S,I,J)=(1,1/2,0)$ 
for our calculation (real part: solid curve, imaginary part: dashed curve).  The 
corresponding threshold and the lowest coupled-channel threshold are represented by the black and red dotted lines, respectively.}\label{fig:plots1}
\end{figure}

\begin{figure}[tbh]
\vspace{8mm}
\begin{center}
\includegraphics[width=1.0\textwidth]{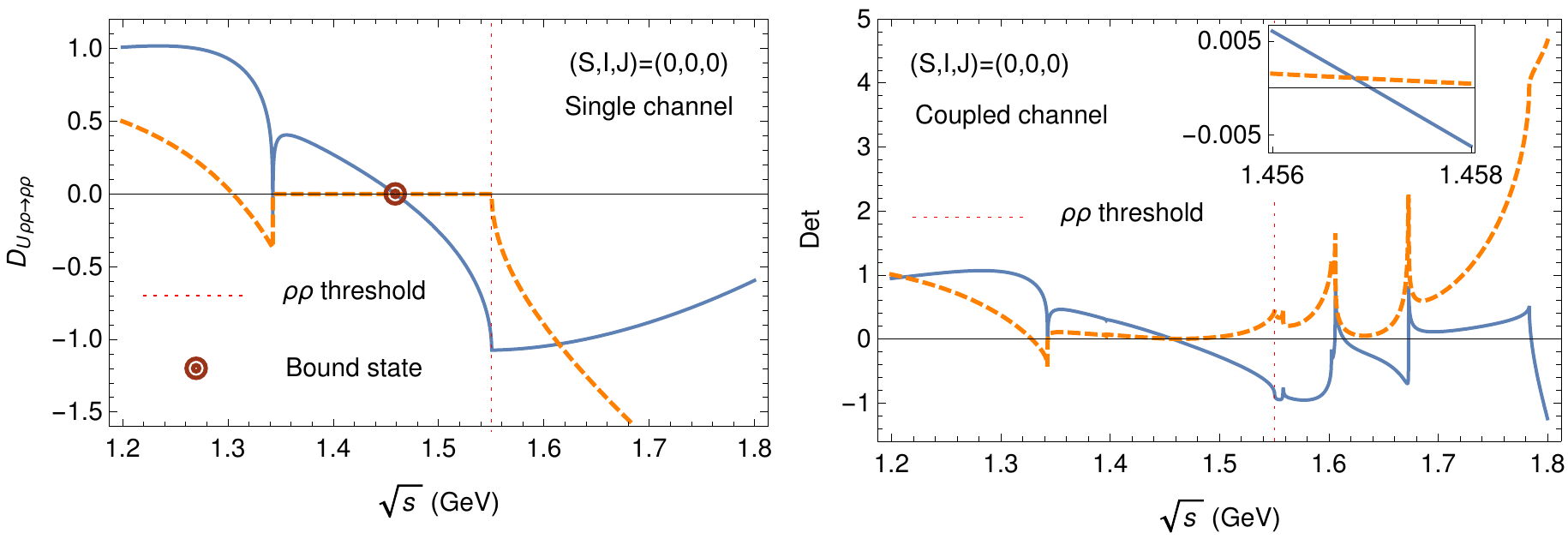}
\end{center}
\vspace{-3mm}
\caption{The $D_U$ function and the determinant defined in Eq.~\eqref{eq:det} of the $(S,I,J)=(0,0,0)$ unitarized amplitudes in
Eq.~\eqref{eq:det} for the single channel  ($\rho\rho$) and the coupled-channel cases evaluated with $q_\text{max}=0.875$~GeV
(real part: solid curve, imaginary part: dashed curve). Note that the imaginary part 
and real part of the determinant for the coupled-channel case do not vanish at the 
same point along the real axis below the threshold, and thus no bound state is 
produced.}\label{fig:plotxbse1}
\end{figure}

As already mentioned in the Introduction, the left-hand cuts of the coupled-channel $T$-matrix in Eq.~\eqref{eq:bse} using the on-shell
factorization are unphysical. 
The locations of the left-hand branch points 
can be easily calculated by making use of Eqs.~\eqref{eq:pwpt} and~\eqref{eq:amps}. For the case of $(S,I)=(0,0)$, while the lowest $s$-channel threshold is 
located at $2M_\rho= 1.550$~GeV and the corresponding right-hand cut runs from $2M_\rho$ to $+\infty$, the left-hand cuts start from 1.606~GeV and
1.602~GeV due to the $t$-channel $\rho$- and $\omega$-exchange diagrams, respectively, for the intermediate process $K^\ast 
\bar{K}^\ast \to K^\ast \bar{K}^\ast$. The two branch points can be clearly seen in the left panel of  Fig.~\ref{fig:plots1}, 
where we plot the potential for the $K^\ast \bar{K}^\ast \to K^\ast \bar{K}^\ast$ employing $g={M_V}/{(2F_\pi)}=4.596$ evaluated using
the average mass of vectors and $F_\pi=93$~MeV.
In addition, there are left-hand branch points at 1.558~GeV and 1.673~GeV due to the $K^*$-exchange
for the processes $K^\ast \bar{K}^\ast\to \omega\phi$ and $\phi\phi$, respectively, which 
are above the $\rho\rho$ threshold as well. The overlapping of the left-hand and the 
right-hand cuts causes a violation of unitarity using the on-shell factorized coupled-channel $T$-matrix in
Eq.~\eqref{eq:bse}, and thus the reliability of the associated poles becomes problematic. A naive use of Eq.~\eqref{eq:bse} 
to the coupled-channel case leads to a branch cut along the whole real axis. The 
pole found in the single-channel case below the $\rho\rho$ threshold~\cite{Gulmez:2016scm} is not reproduced 
in the coupled-channel case, see in Fig.~\ref{fig:plotxbse1}, where the 
determinant Det defined in Eq.~\eqref{eq:det} for the single $\rho\rho$ channel and the coupled-channel cases are 
plotted with the cutoff $q_\text{max}=0.875$~GeV. It confirms the 
invalidation of Eq.~\eqref{eq:bse} for the coupled-channel case with overlapping  
left-hand and right-hand cuts.

The overlapping of the left-hand and the right-hand cuts is present for all the 
coupled-channel cases of $VV$ scattering, i.e. for $(S,I)=(0,0)$, $(0,1)$ and $(1,1/2)$, see, e.g., Fig.~\ref{fig:plots1}.
For $(S,I)=(0,1)$, the left-hand cuts start from 
1.606~GeV and 1.554~GeV for $K^\ast\bar{K}^\ast\to K^\ast\bar{K}^\ast$ (from the $t$-channel $\rho$-exchange) and 
$K^\ast\bar{K}^\ast\to\rho\phi$  (from the $u$-channel $K^*$-exchange), respectively, which are located above the 
lowest threshold $2M_\rho$. The lowest threshold for the 
channel $(S,I)=(1,1/2)$ is 1.667~GeV, while the left-hand branch point of 
$K^\ast \phi\to K^\ast\phi$ from the $u$-channel $K^*$-exchange is located at 1.689~GeV. It is worth mentioning, 
however, that  not all left-hand cuts are present for all partial-wave projected amplitudes. 

For certain values of $J$, the partial wave amplitude vanishes. In 
particular, for $(S,I,J)=(0,0,1)$, the only nonvanishing potential is 
for $ K^\ast \bar{K}^\ast \to  K^\ast \bar{K}^\ast$, such that it reduces to a 
single-channel elastic scattering problem with the normal two-body threshold at $1.783$~GeV, above the corresponding left-hand 
cuts.  In fact, it is easy to show that for all elastic (single-channel) scattering processes, for $s$ in the physical region $s\geq(M_1+M_2)^2$,
we always have $t\leq 0$ and $u\leq (M_1-M_2)^2$. 
For the vector-vector scattering in question, $|M_1-M_2|$ is always smaller than any vector-meson mass, and neither the $t$-channel
nor the $u$-channel exchanged vector meson can go on shell in the scattering physical region. Therefore, the left-hand cut cannot
overlap with the right-hand cut, and the formula in Eq.~\eqref{eq:bse} may still be employed 
in this case.

The right-hand cuts divide the whole energy plane into Riemann sheets. Since the right-hand cuts are included in the loop functions $g_i(s)$, we can
focus on the Riemann sheets of $g_i(s)$. 
Each loop function $g_i(s)$ has two Riemann sheets: the first (physical) and the second (unphysical) sheets, 
denoted as $g_i^\text{I}(s)$ and $g_i^\text{II}(s)$, respectively. In the first sheet, the imaginary part of the center-of-mass momentum
in the corresponding channel is positive, and it is negative in the second sheet.  The expressions of the loop function in the physical
sheet are given in Eqs.~\eqref{eq:ga} and~\eqref{eq:gcut}, while the expressions in the second sheet can 
be obtained by an analytic continuation via~\cite{Oller:1997ti}
\bea
g_i^\text{II}(s)=g_i^\text{I}(s) + 2i\, \rho_i(s) ,
\label{eq:gii}
\eea
with $\rho_i(s)=\sigma_i(s)/(16\pi s)$ the two-body phase space factor.
In this work, we are only interested in the pole structures of the unitarized amplitudes. 
The poles located below the threshold on the real axis in the physical Riemann sheet correspond to possible bound states,
and those on the unphysical Riemann sheets correspond to possible resonances (or virtual states if the poles are on the real axis).

\begin{figure}[tb]
\vspace{8mm}
\begin{center}
\includegraphics[width=1.0\textwidth]{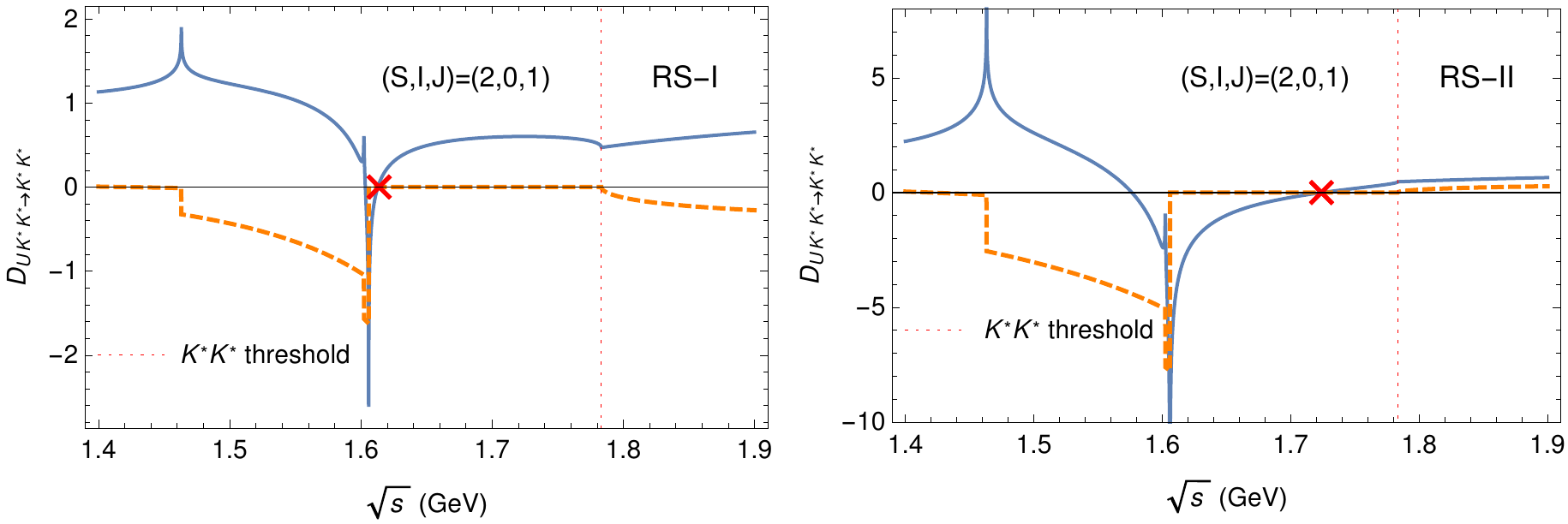}
\end{center}
\vspace{-3mm}
\caption{The $D_U$ functions (real part: solid curve, imaginary part: dashed curve) of $K^\ast K^\ast \to K^\ast K^\ast$ with $(S,I,J)=(2,0,1)$ on the
real axis of physical sheet (RS-I) and unphysical sheet (RS-II). The 
artificial pole due to the unphysical left-hand cut is denoted by the red cross. }\label{fig:plotbse201xRS}
\end{figure}

In this section, we focus on the single-channel cases. According to Bose symmetry, the allowed single-channel $S$-wave vector-meson pairs
include $\rho\rho$ with $(S,I,J)=(0,2,0)$ and $(0,2,2)$, $\rho K^*$ with $(1,3/2,J)$ with $J=0,1$ and 2, $K^*K^*$ with $(2,0,1)$, $(2,1,0)$
and $(2,1,2)$, $K^\ast\bar{K}^\ast$ with $(0,0,1)$.\footnote{For the $S$-wave, 
the $C$ parity of the $K^\ast\bar{K}^\ast$ system with $(S,I,J)=(0,0,1)$ is negative. The system does not couple to the $\omega\phi$,
whose quantum numbers are $(S,I,J)=(0,0,1)$ as well, since the $C$ parity of the latter must be positive.} 
For the $(2,0,1)$ $K^*K^*$ system, a pole located at 1.613~GeV (calculated  with the cutoff value $q_\text{max}=0.875$~GeV), lower than the 
$K^\ast K^\ast$ threshold $1.783$~GeV, is found on the physical Riemann sheet, 
see the left panel of Fig.~\ref{fig:plotbse201xRS}. It might be naively expected to be associated 
with a bound state of $K^\ast K^\ast$. However, in Fig.~\ref{fig:plotbse201xRS},
it is easy to see that the zero, denoted by the red cross, of the determinant which corresponds to a pole of the amplitude is 
caused by the sharp dip due to the presence of the unphysical left-hand cut, starting from 1.606~GeV, which is very close to the pole position. 
In this respect, this pole is just an artifact of  the on-shell factorization in the unitarization formula~\eqref{eq:bse}, and should be absent 
when the on-shell factorization is removed. 
Furthermore, a pole at 1.724~GeV is found on the second Riemann sheet. 
However, from the Fig.~\ref{fig:plotbse201xRS}, it is hard to conclude whether it 
is a dynamically generated state or may also be just a remnant of the 
unphysical left-hand cut. This issue will be revisited by making use of an improved 
unitarization method in the next section.

\begin{figure}[tb]
\vspace{8mm}
\begin{center}
\includegraphics[width=1.0\textwidth]{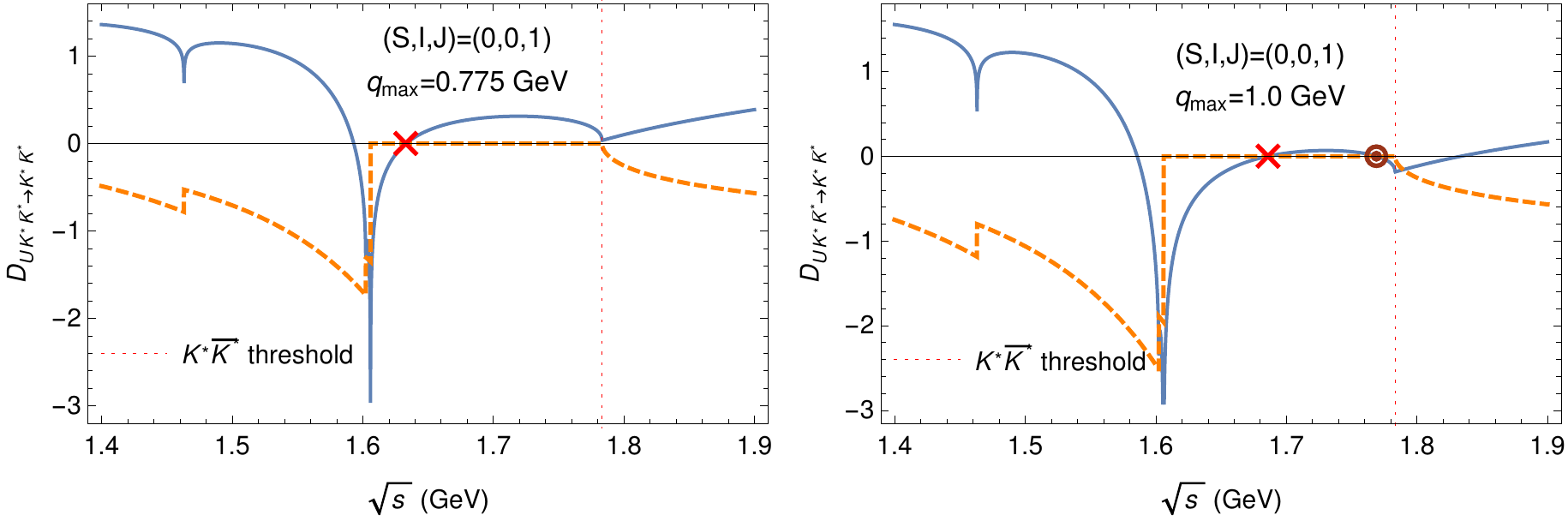}
\end{center}
\vspace{-3mm}
\caption{The $D_U$ functions for the $K^\ast \bar{K}^\ast \to K^\ast \bar{K}^\ast$ with $(S,I,J)=(0,0,1)$ on the real axis on the physical Riemann
sheet with the cutoff $q_\text{max}=0.775$~GeV and 1.0~GeV, respectively. The artificial pole is denoted as red cross and the pole 
associated with a bound state is denoted as a circle.
}\label{fig:plotbse001x}
\end{figure}

\begin{figure}[tb]
\vspace{8mm}
\begin{center}
\includegraphics[width=1.0\textwidth]{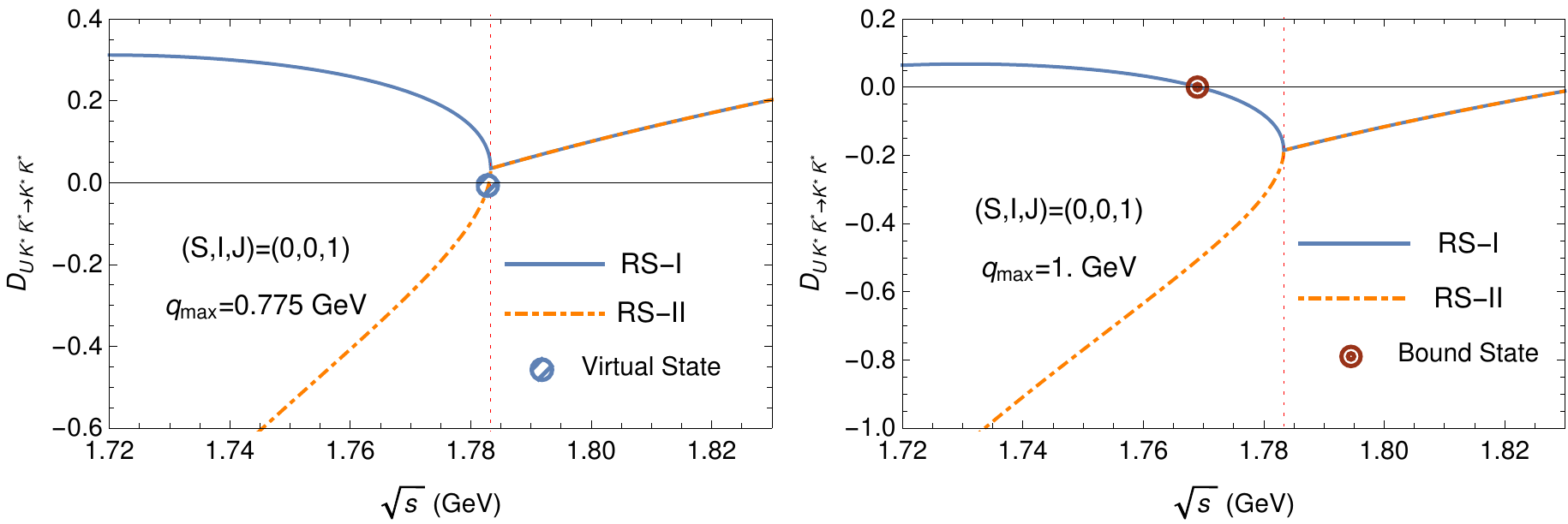}
\end{center}
\vspace{-3mm}
\caption{The real part of the $D_U$ function for $K^\ast \bar{K}^\ast \to K^\ast \bar{K}^\ast$ with $(S,I,J)=(0,0,1)$ on the real axis on the
first (solid line) and second Riemann sheet (dashed line) with the cutoff $q_\text{max}=0.775$~GeV and 1.0 GeV, respectively.}
\label{fig:plotbse001xRS2}
\end{figure}

A similar situation can be found in the $K^*\bar K^*$ system with $(S,I,J)=(0,0,1)$ as well, see 
Fig.~\ref{fig:plotbse001x}, with the left and right panels plotted with $q_\text{max}=0.775$ and 1.0~GeV, respectively. 
Besides a pole near the unphysical left-hand cut and thus artificial, an additional pole close to the threshold appears in the right
panel when $q_\text{max}=1.0$~GeV. 
Because it is close to the threshold, its position and even its presence is sensitive 
to the cutoff value. As the cutoff increases, the pole position 
moves deeper below the threshold, while if we decrease the cutoff the pole 
moves towards the threshold, and at $q_\text{max}=0.808$~GeV the pole disappears 
from the first Riemann sheet and shows up on the second sheet as a virtual state (on the real axis below the threshold). 
The parameter-sensitivity of this pole is shown in Fig.~\ref{fig:plotbse001xRS2} and Table~\ref{tab:boundvirtual}. With different 
choices of the cutoff, the pole appears either as a bound state or a virtual 
state near the threshold. The dependence of the pole on the coupling constant 
$g$ is motivated by its sensitivity to the cutoff, and the values of 4.596 and 4.168 come from using the SU(3)-averaged vector-meson
mass and the $\rho$-meson mass as $M_V$ in $g=M_V/(2F_\pi)$. For $g=4.168$, the pole will become a bound state for $q_\text{max}\geq 1.033$~GeV. 

Apart from the poles located close to the unphysical left-hand cuts and thus are ruled out, the results 
for the single-channel cases in this work are consistent with those in Ref.~\cite{Geng:2008gx} where a pole at $1.802-0.078i$~GeV is reported
corresponding to the pole in Table~\ref{tab:boundvirtual}.~\footnote{Notice that 
the pole has an imaginary part in Ref.~\cite{Geng:2008gx} because the width of the $K^*$ is taken into account  by convolving the
loop integral with the $K^*$ spectral function.} 
This is expected since this pole is very close to the corresponding threshold so that the extreme nonrelativistic approximation in
Ref.~\cite{Geng:2008gx} should work nicely. No pole is found in the other single-channel cases, consistent with the nonrelativistic
results in Ref.~\cite{Geng:2008gx}.

\begin{table}[tb]
\caption{Pole positions (in GeV) evaluated with different values for the cutoff $q_\text{max}$ (in GeV) and the coupling constant $g$.
The subscripts B and V stand for bound state and virtual state, respectively. }\label{tab:boundvirtual}
\vspace{-0.5cm}
\bea
\begin{array}{|c|c|c|c|}
\hline\hline
(S,I,J)=(0,0,1) & q_\text{max}=0.775 & q_\text{max}=0.875 & q_\text{max}=1.0 \\
\hline 
g=4.596 & 1.783_\text{V}  & 1.782_\text{B} & 1.769_\text{B}  \\
\hline
g=4.168 & 1.774_\text{V} & 1.779_\text{V} & 1.783_\text{V} \\
\hline\hline
\end{array}\nonumber
\eea
\end{table}

\section{First iterated solution of the {\boldmath$N/D$} method}\label{sec:nd}
 
In general, the overlapping of the unphysical left-hand cut and the right-hand cut 
breaks unitarity and real analyticity, and thus invalidates the use of the 
formula in Eq.~\eqref{eq:bse}. One way to avoid this problem is to employ 
the $N/D$ dispersion relations~\cite{Chew:1960iv,Bjorken:1960zz,Oller:1998zr,Gulmez:2016scm}. According to the $N/D$ method, a partial wave 
amplitude $T(s)$ can be expressed as a quotient of two functions 
\bea\label{es:nd}
T(s) ={N(s)}{D^{-1}(s)},
\eea
where the denominator $D(s)$ has only the unitary (right-hand) cuts and all the left-hand cuts
are encoded in the numerator $N(s)$. Poles of $T(s)$ correspond to the zeros of the 
$D(s)$ function, which is free of any left-hand cut. For the $S$ wave, one has 
\bea\label{eqs:nd}
D(s) \al = \al \sum_{m=0}^{n-1}\bar{a}_m s^m +\dfrac{(s-s_0)^n}{\pi}
\int_{s_\text{thr}}^\infty ds^\prime \dfrac{\rho(s^\prime)N(s^\prime)}{(s^\prime-s)(s^\prime -s_0)^n} ,\nonumber \\
N(s) \al = \al \sum_{m=0}^{n-1}\bar{a}_m^\prime s^m + \dfrac{(s-s_0)^n}{\pi}
\int_{-\infty}^{s_\text{left}}ds^\prime \dfrac{\text{Im} T(s^\prime)D(s^\prime)}{(s^\prime -s_0)^n(s^\prime -s)},
\eea
where the so-called Castiliejo--Dalitz--Dyson (CDD) poles~\cite{Castillejo:1955ed}
are not included, and $n$ is the number of subtractions 
needed to ensure the convergence such that 
\bea
\lim_{s\to \infty}\dfrac{N(s)}{s^n}=0.
\eea
Eq.~\eqref{eqs:nd} constitutes a system of integral equations in which an input 
given by $\text{Im}T(s)$ along the left-hand cut is needed. 

 Instead of giving a complete treatment of the $N/D$ method, see e.g. in Refs.~\cite{Noyes1960:1736,Guo:2013rpa}, in this section we approximate the 
$N(s)$ function by the tree-level amplitudes $V(s)$. This leads to the first iterated 
solution as proposed in Ref.~\cite{Gulmez:2016scm}. Then we have
\bea\label{eqs:ndc}
N(s)_{ij}\al = \al V(s)_{ij},\\
D(s)_{ij}\al = \al \gamma_{0 ij} +\gamma_{1ij} (s-s^i_\text{thr})+\frac{1}{2}\gamma_{2ij}(s-s^i_\text{thr})^2
+\dfrac{(s-s^i_\text{thr})s^2}{\pi}\int_{s_\text{thr}^i}^\infty 
ds^\prime \dfrac{\rho(s^\prime)_iV(s^\prime)_{ij}}{(s^\prime-s^i_{\text{thr}})(s^\prime 
-s)s^{\prime 2}} , \nonumber
\eea
with the $\gamma$ parameters the subtraction constants and the subscripts $i$ and $j$ the 
channel indices. 
One notices that the potential matrix $V(s)$ is inside the dispersion integral in $D(s)$.
By construction, the $D(s)$ matrix elements 
in Eq.~\eqref{eqs:ndc} are free of left-hand cuts and thus unitarity is ensured by using Eqs.~\eqref{es:nd} and \eqref{eqs:ndc}. 
As we have shown in the paragraph before Eq.~\eqref{eq:gii} in section~\ref{sec:bse}, for each individual $VV$ scattering process, the left-hand branch points for $V_{ij}$ from the $t$- and $u$-channel vector-exchange diagrams  are below the $s$-channel
thresholds for both $i$ and $j$ channels. As a consequence, the unknown subtraction constants in the $D(s)$ functions 
may be fixed by matching to the denominator of Eq.~\eqref{eq:bse} around 
the thresholds, that is~\cite{Gulmez:2016scm} 
\bea\label{eq:match}
\gamma_{0ij}+\gamma_{1ij}(s-s^i_\text{thr})+\frac{1}{2}\gamma_{2ij}(s-s^i_\text{thr})^2 \al 
= \al \delta_{ij} -g_i(s)V(s)_{ij} \nonumber\\ \al - \al \frac{(s-s^i_\text{thr})s^2}{\pi}\int_{s^i_\text{thr}}^\infty ds^\prime
\frac{\rho(s^\prime)_iV(s^\prime)_{ij}}{(s^\prime-s^i_\text{thr})(s^\prime -s) s^{\prime 2} } \nonumber\\
\al \equiv \al w(s)_{ij},
\eea
where contributions of $\mathcal{O}\big((s-s_\text{thr})^3\big)$ are neglected. The  
subtraction constants depend on the cutoff $q_\text{max}$ via the matching 
conditions \eqref{eq:match}. 
In this way, one has 
\bea
\gamma_{0ij} \al =\al w(s_\text{thr})_{ij}, \nonumber\\
\gamma_{1ij} \al =\al   w^{\prime}(s^i_\text{thr})_{ij},
\nonumber\\
\gamma_{2ij} \al =\al   w^{\prime\prime }(s^i_\text{thr})_{ij}.
\eea
The so-obtained $N$ and $D$ functions have the correct analytical properties with 
appropriate cuts.

\subsection{Single-channel cases}\label{sec:nd1}

\begin{figure}[tb]
\begin{center}
\includegraphics[width=1.\textwidth]{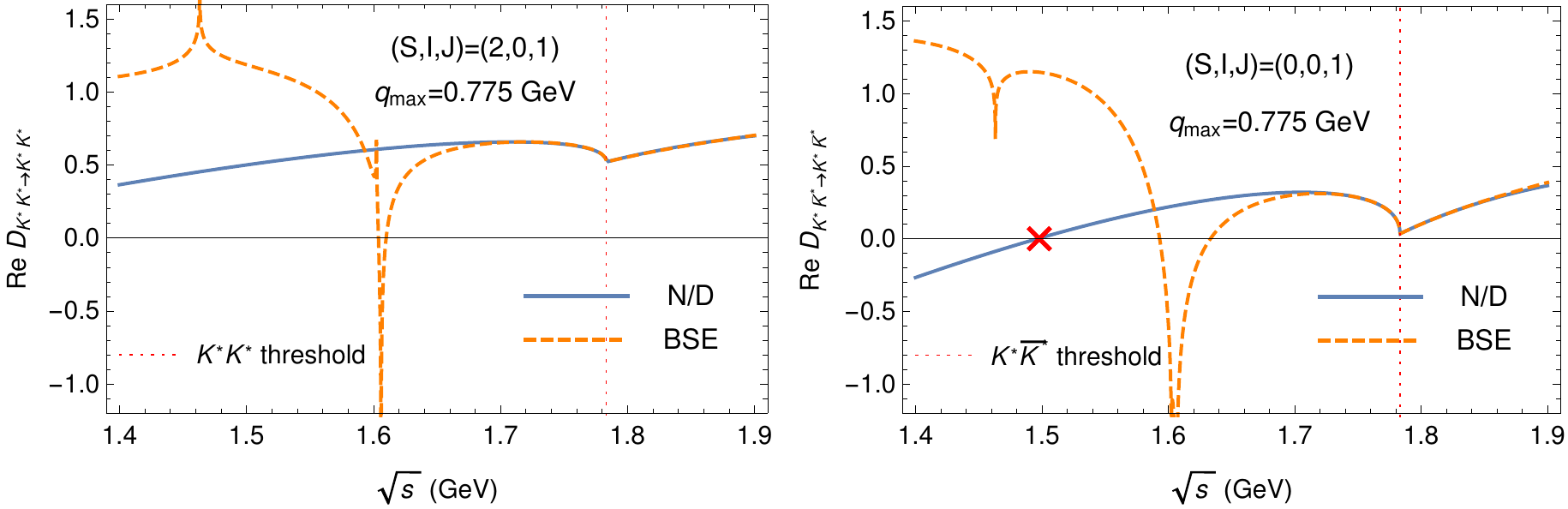}
\end{center}
\vspace{-3mm}
\caption{Real parts of the $D(s)$ functions for the $(0,0,1)$ and $(2,0,1)$ cases based on 
the $N/D$ method and the on-shell factorization for the BSE. The pole 
due to the left-hand cut is absent for $(2,0,1)$, and is moved to very deep  below the threshold (around 1.49~GeV)
where the matching \eqref{eq:match} is in question.}\label{fig:plotsinglechannel}
\vspace{-5mm}
\end{figure}

In this subsection, we consider poles of the scattering amplitudes for the single-channel cases. In order to consider the possible
poles on the unphysical  Riemann sheet, the following continuation of the scattering amplitude is employed:
\bea \label{eq:usualCont}
T^{\text{II}}(s) = \frac{1}{[T^\text{I}(s)]^{-1}-2i\rho}
\eea 
with $T^\text{I}(s)=N(s)/D(s)$ the amplitude in the physical Riemann sheet as given in Eq.~\eqref{es:nd}.
Since the $D(s)$ function, i.e. Eq.~\eqref{eq:match}, is free of unphysical cuts, it is expected that the poles due to the
presence of unphysical left-hand cuts found using the on-shell factorization
should be absent in the $N/D$ method. 
This is indeed the case for the pole on the first Riemann sheet in $K^*K^*$ scattering, whose quantum numbers are $(S,I,J)=(2,0,1)$,
as can be seen by comparing the left panel of Fig.~\ref{fig:plotbse201xRS} and the left panel of
Fig.~\ref{fig:plotsinglechannel}. For the $K^*\bar K^*$ scattering with $(S,I,J)=(0,0,1)$, the pole on the first 
Riemann sheet is found much deeper than that in the BSE, see the right panel of Fig.~\ref{fig:plotsinglechannel}.
However,
the pole is located far away from the $K^*\bar K^*$ threshold such that 
interactions among the vector mesons other than those considered here should become relevant, and thus such a pole is not reliable.
This is reflected by the dramatic dependence of the pole position on the coupling and the cutoff values:
it is located between 1.42~GeV and 1.64~GeV by varying the cutoff $q_\text{max}$ from 0.775~GeV to 1.0~GeV and the coupling $g$ from 
4.168 to 4.596.
In contrast, the pole  on the second Riemann sheet for the $K^*K^*$ with $(S,I,J)=(2,0,1)$, being much closer to the $K^*K^*$ threshold,
changes only a little with the variation of the coupling constant and the cutoff, and it is a virtual state pole on the real axis.
Finally, the near-threshold pole for the $K^*\bar K^*$ with $(S,I,J)=(0,0,1)$ corresponding to either a bound state or a virtual state,
depending on the parameter values, changes 
barely in the two unitarization procedures. It is not a surprise since it is located far away from the unphysical left-hand cut in
the on-shell factorization method and the difference between the two methods is of $\mathcal{O}\big((s-s_\text{thr})^3\big)$.
Similar to the on-shell factorization of the BSE case, 
no other poles are found for the single-channel cases.

\subsection{Coupled-channel cases}\label{sec:nd2}

In this subsection, we investigate poles of the $VV$ scattering amplitudes in the coupled-channel cases using the $N/D$ 
method. For the $n$-channel case, there are 
$2^n$ Riemann sheets ($1$ physical Riemann sheet and $2^n-1$ unphysical ones) in total. In addition to the bound state poles
on the physical sheet, poles may also appear on the unphysical ones. The continuation to the unphysical Riemann sheets for coupled-channels is given by the matrix form of Eq.~\eqref{eq:usualCont}, where the $\rho(s)$ is 
a diagonal matrix $\rho(s)=\text{diag}\{N_i\rho_i(s)\}$ with $i$ the channel index and $N_i=0$ and $1$ representing the physical and unphysical sheets for the 
$i^\text{th}$ channel. It is easy to see that the $2^n$ Riemann sheets correspond to the various choices of the set $N_i$. In this section,  we only consider the poles which have significant impact
on the physical observables in a specific energy region. 
In Table~\ref{tab:poles}, we collect the poles in the energy region of interest
for the coupled-channel calculations using different values of coupling $g$ and
cutoff $q_\text{max}$. Nevertheless, it is possible to find unphysical poles whose distance to the relevant thresholds is
larger than $250~\mathrm{MeV}$ with an imaginary part larger than 200~MeV on the physical sheet for $J=2$ sectors
due to the perturbative treatment of the $N(s)$ functions. Such poles exist as long as there is no bound state pole, as required for a nontrivial holomorphic function.\footnote{They are not harmful as long as they are located far from the relevant energy region (beyond the applicable range of the theory).} 
This kind of poles does not appear for $J=0$ and $J=1$ sectors.  
This is due to the fact that the contribution from the on-shell approximation of exchanged diagrams to the $N$ functions for $J=2$ sectors
are significantly larger than those for $J=0$ and $J=1$ sectors,
see e.g. Fig.~4 in Ref.~\cite{Gulmez:2016scm}. Notice that any pole in the physical Riemann sheet other than the bound state ones
violates causality (see, e.g.,~\cite{Gribov:2009zz}).
To overcome such an unsatisfactory situation, without knowing exactly the discontinuity along the left-hand cuts, so that they
can be treated nonperturbatively,\footnote{Recently, the left-hand cuts are treated nonperturbatively in the $N/D$ method for the
$S$-wave nucleon-nucleon scattering in Ref.~\cite{Entem:2016ipb}.} we exclude those poles with an imaginary
part more than $200~\mathrm{MeV}$ and the distance to each threshold more than $250~\mathrm{MeV}$ by
hand as done in Refs.~\cite{Borasoy:2006sr,Mai:2012dt}.

\begin{table}[tb]
\caption{Poles (in units of GeV) found in different channels using different values of the coupling $g$ and cutoff $q_\text{max}$.
Here, $g_1=4.596$ and $g_2=4.168$, and $q_\text{max1,2,3}=
0.775$, 0.875 and 1.0~GeV, respectively. Only the poles located on the physical 
sheet and those on unphysical sheets with a significant impact on physical
observables are listed. 
The dominant channels (DCs) are listed in the second column. The notation follows that of Table~\ref{tab:boundvirtual}.}\label{tab:poles}
\vspace{-0.5cm}
\bea
\begin{array}{|c|c|c|c|c|c|c|c|}
\hline\hline 
(S,I,J) & \text{DC} & g_1,q_\text{max1} & g_1,q_\text{max2} & 
g_1,q_\text{max3} & g_2,q_\text{max1} & g_2,q_\text{max2} & 
g_2,q_\text{max3}\\
\hline 
(0,0,0) & \rho\rho & 1.47_\text{B} & 1.44_\text{B} & 1.41_\text{B} & 1.50_\text{B} & 1.48_\text{B} & 1.45_\text{B} \\
\, & K^\ast \bar{K}^\ast & 1.69\pm 0.02i & 1.64\pm 0.02i & 1.58\pm 0.02i & 1.76+0.01i & 1.73\pm 0.02i & 1.67\pm 0.02i \\
\hline 
(0,1,0) & K^\ast \bar{K}^\ast & 1.79\pm 0.01i  & 1.78 \pm 0.02i & 1.76 \pm 0.03i   & 1.79\pm 0.01i & 1.77 \pm 0.02i & 1.75 \pm 0.03i \\
\hline 
(0,1,1) & \rho\rho & 1.47_\text{V} & 1.48_\text{V} & 1.50_\text{V} & 1.44_\text{V} & 1.45_\text{V} & 1.46_\text{V} \\
\hline 
(1,1/2,0) & \rho K^\ast & 1.64_\text{B} & 1.62_\text{B} & 1.58_\text{B} & 1.66_\text{B} & 1.65_\text{B} & 1.63_\text{B} \\
\hline\hline
\end{array}\nonumber
\eea
\end{table}

\begin{figure}[tb]
\vspace{8mm}
\begin{center}
\includegraphics[width=1.\textwidth]{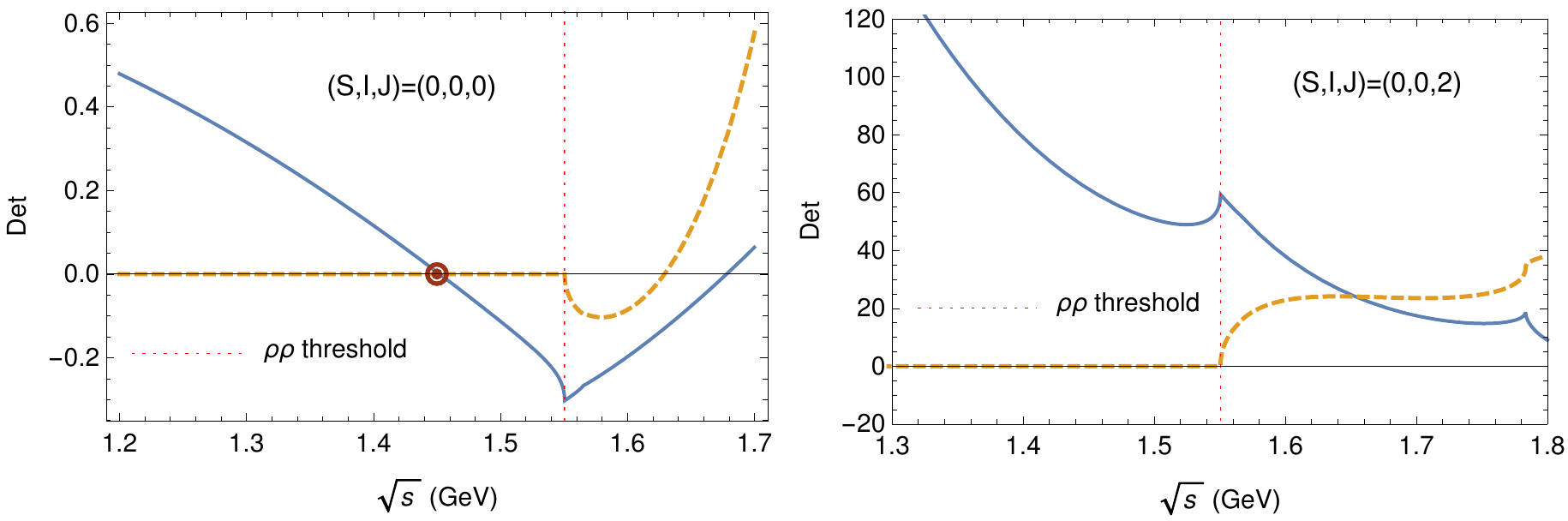}
\end{center}
\vspace{-3mm}
\caption{Determinant of the $D(s)$ matrix for the cases with $(S,I,J)=(0,0,0)$ and 
$(0,0,2)$ on the first Riemann sheet evaluated using $q_\text{max}=0.875$~GeV (real part: solid curve, imaginary part: dashed curve).
}\label{fig:plotnd00J875g1x}
\end{figure}

A comparison of the poles obtained in this work with those in Ref.~\cite{Geng:2008gx} using the nonrelativistic limit is
presented in Table~\ref{tab:summary}, together with the possible assignments to the mesons listed in the Review of Particle Physics
by the Particle Data Group (PDG)~\cite{pdg2018}. We find a bound state pole in the $(S,I,J)=(0,0,0)$ system, which is absent in the on-shell factorization method,  and it is
shown in the left panel of Fig.~\ref{fig:plotnd00J875g1x}. 
The pole is located between 1.4 and 1.5~GeV, and it might be significant to form the scalar meson $f_0(1370)$ or $f_0(1500)$~\cite{pdg2018}.
It is consistent with the pole found in the SU(2) case~\cite{Gulmez:2016scm} and in the 
nonrelativistic limit in Ref.~\cite{Geng:2008gx}, see Table~\ref{tab:summary}. One may wonder about the reliability of this pole since it
is about 0.6~GeV below the $\phi\phi$ threshold while the subtraction constants in $D(s)_{55}$, where $5$ labels the $\phi\phi$ channel,
are fixed by matching at the $\phi\phi$ threshold via Eq.~\eqref{eq:match}. To investigate this issue, we calculate its residues which 
correspond to the products of its couplings to various channels, i.e.,
\bea
 {g_ig_j} = \lim_{s\to s_\text{pole}} (s-s_\text{pole})\, T_{ij}(s),
\eea
with $i$ the channel index and $g_i$ the corresponding effective coupling constant. 
It turns out that this bound state pole couples most strongly to the $\rho\rho$ 
channel (dominant channel) and the channels with higher thresholds are negligible. 
In addition, using $g=4.618$ and $q_\text{max}=0.875$~GeV a pole located at $(1.73\pm 0.02\,i)$~GeV is found on an unphysical 
Riemann sheet. 
In Ref.~\cite{Geng:2008gx}, the corresponding pole is at $(1.726\pm 0.028\,i)$~GeV, 
and is assigned to the $f_0(1710)$ scalar meson. 
It is not surprising  that the two pole positions are similar since the threshold of the dominant channel $K^\ast\bar{K}^\ast$
is close to the pole, and thus the nonrelativistic limit in Ref.~\cite{Geng:2008gx} is a good approximation. Nevertheless, we find a
sizable dependence of the pole position on the coupling and cutoff values, see Table~\ref{tab:poles}.

\begin{table}[tb]
\caption{Comparison of the poles obtained in this work with those in Ref.~\cite{Geng:2008gx}. Only the real parts of the pole positions
(in GeV) are given. The ranges of our results cover those obtained using the parameters in Table~\ref{tab:poles}.
Possible assignments to the known mesons are also listed together with their mass ranges given by the PDG~\cite{pdg2018}.
}\label{tab:summary}
\vspace{-0.5cm}
\bea
\begin{array}{|l|c|c|c|c|c|}
\hline\hline
S &I^G(J^{PC}) & \text{Pole}~[\text{GeV}] & \text{PDG} & \text{Mass}~[\text{GeV}]~\text{\cite{pdg2018}} & \text{Pole}  ~[\text{GeV}]~\text{\cite{Geng:2008gx}}\\
\hline
\multirow{5}{*}{0} & 0^+(0^{++}) & [1.41-1.50] & f_0(1370) & [1.2-1.5] &  1.512 \\
\, &             &             & f_0(1500) & 1.504\pm0.006 & \\
\, & 0^+(0^{++}) & [1.58-1.76] & f_0(1710) & 1.723^{+0.006}_{-0.005} & 1.726 \\
\, & 0^-(1^{+-}) & [1.77-1.78] & - & - & 1.802 \\
\, & 1^-(0^{++}) & [1.75-1.79] & a_0(1950)? & 1.931\pm0.026 & 1.780 \\
\, & 1^+(1^{+-}) & [1.44-1.50] & - & - & 1.679 \\
\hline
\multirow{1}{*}{1} & 1/2(0^+) & [1.58-1.66] & - & - & 1.643\\
\hline
2 & 0({1^+}) & [1.68-1.74] & - & - & - \\
\hline\hline
\end{array}\nonumber
\eea
\end{table}

The poles we obtained in the scalar and vector sectors of $(S,I)=(0,0)$ are consistent with 
the nonrelativistic results in Ref.~\cite{Geng:2008gx}, see Table~\ref{tab:summary}.  In Ref.~\cite{Geng:2008gx}, two bound state poles are
reported in the tensor sector, which couple dominantly to $\rho\rho$ and $K^\ast \bar{K}^\ast$ and are assigned to the $f_2(1270)$ and
$f_2^\prime (1525)$ resonances, respectively.\footnote{Notice that at 1.27~GeV the binding momentum of the $\rho\rho$ system is around 0.44~GeV,
and at 1.53~GeV the binding momentum of the $K^*\bar K^*$ system is around 0.45~GeV.} However, no such poles are found in our calculation in
this $(0,0,2)$ channel, see the right panel of Fig.~\ref{fig:plotnd00J875g1x}, which agrees with the SU(2) relativistic result in
Ref.~\cite{Gulmez:2016scm}. We also want to point out that although we exclude by hand the remote region on the physical Riemann sheet from
the applicability of our treatment, see discussions at the beginning of this subsection, there is not any trend for the emergence of a bound state
pole all the way from the $\rho\rho$ threshold to around $1270$~MeV, see Fig.~\ref{fig:plotnd00J875g1x}. We thus regard the absence of tensor bound
state poles as a reliable conclusion. To investigate more quantitatively possible poles beyond the near-threshold region, a more rigorous and
complete treatment to the left-hand cuts is required. 
Also, a more realistic vector-vector interaction beyond the leading order hidden local symmetry Lagrangian \eqref{lag:hg} is needed,
which do not seem to  be available in the near future. 

\begin{figure}[tb]
\vspace{8mm}
\begin{center}
\includegraphics[width=1.\textwidth]{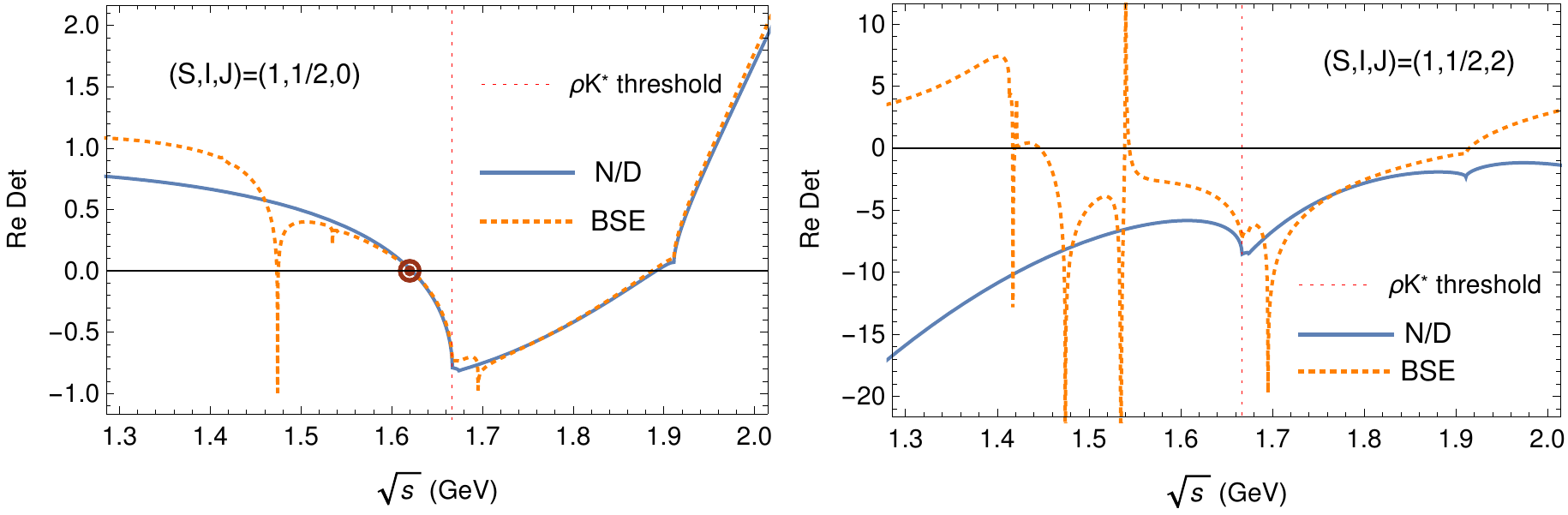}
\end{center}
\vspace{-3mm}
\caption{Real part of the determinant of the $D(s)$ matrix for the cases of $(S,I,J)=(1,1/2,0)$ and 
$(1,1/2,2)$ on the physical Riemann sheet evaluated with $q_\text{max}=0.875$~GeV and $g=4.596$. A bound state is found in the scalar sector,
while no bound state is found in the tensor sector. 
}\label{fig:plotnd1h01}
\end{figure}

For the $(S,I,J)=(1,1/2,0)$ system, a pole around 1.6~GeV coupled dominantly to $\rho K^*$ is found, which corresponds to the bound state pole
at $(1.643\pm 0.047\,i)$~GeV in Ref.~\cite{Geng:2008gx}\footnote{Note that the imaginary part is due to the convolution of the loop integrals
with the spectral functions of the vector mesons in 
Ref.~\cite{Geng:2008gx}.}. However, for the $(S,I,J)=(1,1/2,2)$ tensor sector, we do not find a bound state pole corresponding to the one
at ($1.431\pm 0.001i$)~GeV in Ref.~\cite{Geng:2008gx}, see the right panel of Fig.~\ref{fig:plotnd1h01}.

So far we have neglected the widths of the vector mesons. A convolution 
of the loop integrals with the vector-meson spectral functions, accounting for the contributions from the decays $\rho\to \pi\pi$ and
$K^\ast \to \pi K$, will give a width to the bound states and increase the width of resonances. In addition, box diagrams with four intermediate 
pseudoscalar mesons contribute to the widths as well, corresponding to the decays into 
two pseudoscalar mesons. In order to include such contributions, one 
has to introduce model-dependent form factors, and the widths strongly 
depend on the cutoff one uses in the form factors. However, the real part of 
the pole positions are merely affected~\cite{Gulmez:2016scm,Molina:2008jw,Geng:2008gx}. 
Since we are only interested in the existence of poles, we therefore do not introduce the convolution of loop functions and the box diagrams in this work.

\section{Summary}\label{sec:summary}

In Ref.~\cite{Gulmez:2016scm}, the first-iterated $N/D$ method was proposed to the $\rho\rho$ interaction and possible dynamically generated
resonances, and no tensor bound state poles reported in the nonrelativistic calculation~\cite{Molina:2008jw} were found. In this method,
poles of the $T$-matrix correspond to zeros of the $D(s)$ function, which does not have any left-hand cut. 
Thus, contrary to the claim in Ref.~\cite{Geng:2016pmf}, the disappearance of the tensor bound state poles is not due to the unphysical
left-hand cuts, but due to the energy-dependence of the potential. 

In this paper, we extend the method to coupled channels for the $S$-wave interactions for the whole vector-meson nonet. 
The possible dynamically 
generated resonances (including bound states) are listed in Table~\ref{tab:summary}. Contrary to the results in the nonrelativistic
treatment~\cite{Geng:2008gx},  we did not find any bound state poles in tensor sectors with $I(J^P)=0(2^+)$ and $1/2(2^+)$, confirming
the single-channel result for the $\rho\rho$ case in Ref.~\cite{Gulmez:2016scm}.

In this work, the left-hand cut is considered perturbatively. 
In addition, the contributions from the large widths of the $\rho$ and the $K^\ast$, as well as the channels
with a pair of pseudoscalar mesons, are not considered. The inclusion of such contributions will increase the widths of the generated states.
The results in this work are phenomenologically valuable to understand which mesonic resonances owe their existence mainly due to the
interactions between a pair of vector mesons.

\section*{Acknowledgements}
This work is partially supported
by the National Natural Science Foundation of China (NSFC) and Deutsche Forschungsgemeinschaft (DFG) through 
funds provided to the Sino--German Collaborative Research Center ``Symmetries and the
Emergence of Structure in QCD'' (NSFC Grant No.~11621131001,
DFG Grant No.~TRR110), by the NSFC (Grant No.~11747601), by the Thousand Talents Plan for Young
Professionals, by the CAS Key Research Program of Frontier Sciences
(Grant No.~QYZDB-SSW-SYS013), by the CAS Key Research Program (Grant No.~XDPB09), by the CAS Center for Excellence in Particle Physics (CCEPP), 
by the by the CAS President's International Fellowship Initiative (PIFI) (Grant No.~2018DM0034), 
and by VolkswagenStiftung (Grant No. 93562).

\end{document}